\documentclass[aip,jmp,showkeys,
preprint,
11pt,
]{revtex4-1}

\usepackage{amsmath}
\usepackage{amsthm}
\usepackage{amssymb}
\usepackage{bm}
\usepackage{graphicx}
\usepackage{array}
\usepackage{dsfont}
\usepackage{caption}
\usepackage{dashrule}
\usepackage{mathtools}
\usepackage{MnSymbol}
\usepackage{fancyhdr,color}
\usepackage[hang]{footmisc}
\usepackage{hyperref}
\hypersetup{
citebordercolor= {1 0 0},
linkbordercolor={0 0 1},
}

\newtheorem{theorem}{Theorem}[section]
\newtheorem{corollary}[theorem]{Corollary}

\newtheorem{definition}[theorem]{Definition}
\newtheorem{lemma}[theorem]{Lemma}

 \newcommand{\diagdots}[3][-25]{%
  \rotatebox{#1}{\makebox[0pt]{\makebox[#2]{\xleaders\hbox{$\cdot$\hskip#3}\hfill\kern0pt}}}%
}

\newcommand{\diag}{\mathrm{diag}}

\begin{document}


\title{\texorpdfstring{A new algorithm for computing branching rules and Clebsch--Gordan coefficients of unitary representations of compact groups\,\linespread{1}\footnote{\noindent\;The authors would like to thank the financial support provided by Ministry of Economy and Competitivity of Spain, under the grant MTM2014-54692-P, Community of Madrid research project QUITEMAD+, S2013/ICE-2801, and the Office of Naval Research Global, N62909-15-1-2011.}}{A new algorithm for computing branching rules and Clebsch--Gordan coefficients of unitary representations of compact groups}}


\author{A. Ibort\,}%
\thanks{\noindent\;Electronic mail: \href{mailto:albertoi@math.uc3m.es}{albertoi@math.uc3m.es}}
\affiliation{ Instituto de Ciencias Matem\'aticas (CSIC-UAM-UC3M-UCM) and
Dpto. de Matem\'aticas, Univ. Carlos III de Madrid,  Avda. de la
Universidad 30, 28911 Legan\'es, Madrid, Spain.
}%

\author{A. L\'opez Yela\,}%
\thanks{\noindent\;Electronic mail: \href{mailto:alyela@math.uc3m.es}{alyela@math.uc3m.es}}
\affiliation{Dpto. de teor\'ia de la se\~nal y telecomunicaciones, Univ. Carlos III de Madrid,  Avda. de la
Universidad 30, 28911 Legan\'es, Madrid, Spain.
}%

\author{J. Moro\,}%
\thanks{\noindent\;Electronic mail: \href{mailto:jmoro@math.uc3m.es}{jmoro@math.uc3m.es}}
\affiliation{Dpto. de Matem\'aticas, Univ. Carlos III de Madrid,  Avda. de la
Universidad 30, 28911 Legan\'es, Madrid, Spain.
}%

\begin{abstract}
A numerical algorithm that computes the decomposition of any finite-dimen\-sio\-nal unitary reducible representation of a compact Lie group is presented.  The algorithm, which does not rely on an algebraic insight on the group structure, is inspired by quantum mechanical notions. After generating two adapted states (these objects will be conveniently defined in {\bf Def.\,\ref{adapted_def}}) and after appropriate algebraic manipulations, the algorithm returns the block matrix structure of the representation in terms of its irreducible components.  It also provides an adapted orthonormal basis.    The algorithm can be used to compute the Clebsch--Gordan coefficients of the tensor product of irreducible representations of a given compact Lie group.   The performance of the algorithm is tested on various examples: the decomposition of the regular representation of two finite groups and the computation of Clebsch--Gordan coefficients of two examples of tensor products of representations of  $SU(2)$.
\end{abstract}

\keywords{Clebsch--Gordan coefficients,  Compact groups, Unitary representations, Quantum states, Numerical algorithm}
\pacs{02.20.Qs, 02.60.Dc}

\maketitle



\section{Introduction}\label{sec:introduction}

The algorithm presented in this paper solves the problem of {\textit{numerically}}
determining the decomposition of a finite-dimensional
irreducible unitary linear representation (`irrep' in what
follows) of a compact group $G$ with respect to the unitary irreducible
representations (irreps) of a given subgroup $H \subset G$.

More precisely, let $G$ be a compact Lie group and $(\mathcal{H},
U)$ a  finite-dimensional irreducible unitary representation of
it, i.e., $U \colon G \to U(\mathcal{H})$ is a group homomorphism that satisfies the following three conditions:
\phantomsection\label{Group_conditions_ps}\begin{alignat*}{2}
\hspace{-1.5cm}\begin{array}{c}
\vspace{-1cm}\phantom{a}\\
(C.1)\\
\vspace{-1cm}\phantom{a}
\end{array}& \hspace{2.3cm} \begin{array}{c}
\vspace{-1cm}\phantom{a}\\
U(g_1g_2)=U(g_1)U(g_2),\qquad\text{for all $g_1,g_2\in G$.}\\
\vspace{-1cm}\phantom{a}
\end{array}\\
\hspace{-1.5cm}\begin{array}{c}
\vspace{-1cm}\phantom{a}\\
(C.2)\\
\vspace{-1cm}\phantom{a}
\end{array}&\hspace{2.3cm}\begin{array}{c}
\vspace{-1cm}\phantom{a}\\
U(e)=\mathds{1}\,.\\
\vspace{-1cm}\phantom{a}
\end{array}\\
\hspace{-1.5cm}\begin{array}{c}
\vspace{-1cm}\phantom{a}\\
(C.3)\\
\vspace{-1cm}\phantom{a}
\end{array}&\hspace{2.3cm}\begin{array}{c}
\vspace{-1cm}\phantom{a}\\
U(g^{-1})=U(g)^{-1}=U(g)^\dagger,\qquad\text{for all $g\in G$.}\\
\vspace{-1cm}\phantom{a}
\end{array}
\end{alignat*}
Here, $\mathcal{H}$ is a complex Hilbert space with  inner product $\langle \cdot, \cdot \rangle$,
$U(\mathcal{H})$ is the group of unitary operators on
$\mathcal{H}$ and $^\dagger$ stands for the adjoint.

Conditions \hyperref[Group_conditions_ps]{$(C.1)$\,--\,$(C.3)$} above define a \textit{unitary representation}
$(\mathcal{H},U)$ of the group $G$. The representation is said to
be \textit{irreducible} if there are no proper invariant subspaces of
$\mathcal{H}$, i.e., if any linear subspace $W\subset \mathcal{H}$
is such that $U(g) W \subset W$ for all $g\in G$, then $W$ is
either $\{ \mathbf{0} \}$ or $\mathcal{H}$. Since the group $G$ is
compact, any irreducible representation of $G$ will be
finite-dimensional with dimension say $n$ ($n = \dim \mathcal{H}$).

Consider a closed subgroup $H \subset G$.  The restriction of $U$
 to $H$ will define a unitary representation of $H$ which is reducible in general, that is, it will possess invariant subspaces $\mathcal{L}^\alpha$ such that $U(h)\mathcal{L}^\alpha \subset \mathcal{L}^\alpha$
 for all $h\in H$.   If we denote by $\widehat{H}$ the family of equivalence
 classes of irreps of $H$ (recall that two unitary representations
 of $H$, $V \colon H\to U(E)$ and $V^\prime \colon H \to U(E')$, are
 equivalent if there exists a unitary map $T \colon E \to \displaystyle{E'}$
 such that $\displaystyle{V}'(h) \circ T = T \circ V(h)$ for all $h \in H$), then:
 \begin{equation}\label{decomposition}
 \mathcal{H} = \bigoplus_{\alpha \in \widehat{H}} \mathcal{L}^\alpha \, ,\qquad \mathcal{L}^\alpha=c_\alpha\mathcal{H}^\alpha\, = \bigoplus_{a=1}^{c_ \alpha} \mathcal{H}^\alpha \, ,
\end{equation}
where $c_\alpha$ are non-negative integers, $\{ \alpha \}$ denotes
a subset in the class of irreps  of the group $H$ (each
$\alpha$ denotes a finite-dimensional irrep of $H$ formed by the
pair $(\mathcal{H}^\alpha, U^\alpha)$) and $c_\alpha
\mathcal{H}^\alpha$ denotes the direct sum of the linear space
$\mathcal{H}^\alpha$ with itself $c_\alpha$ times.    Thus,  the
family of non-negative integer numbers $c_\alpha$ denotes the
multiplicity of the irreps $(\mathcal{H}^\alpha,U^\alpha)$ in $(\mathcal{H},U)$.
The numbers $c_\alpha$ satisfy $n = \sum_{\alpha} c_\alpha
n_\alpha$ where $n_\alpha = \dim \mathcal{H}^\alpha$ and the
invariant subspaces $\mathcal{L}^\alpha$ have dimension $c_\alpha
n_\alpha$.  Notice that the unitary operator $U(h)$ will
have the corresponding block structure:
\begin{equation}\label{decompositionU}
U(h) = \bigoplus_{\alpha \in \widehat{H}} c_\alpha U^\alpha(h) \, , \qquad \forall h \in H \, ,
\end{equation}
where $\displaystyle{U}^\alpha (h ) = U(h) \mid_{\mathcal{H}^\alpha} $.

The problem of determining an orthonormal basis of $\mathcal{H}$
adapted to the decomposition (\ref{decomposition}) will be called
the {\em Clebsch--Gordan problem} of $(\mathcal{H},U)$ with respect to
the subgroup $H$.  To be more precise, the Clebsch--Gordan problem
of the representation $U$ of $G$ in $\mathcal{H}$ with respect to
the subgroup $H$ consists in finding an orthonormal basis $\{
u_{a,k}^\alpha \mid \alpha \in \widehat{H}, a = 1,\ldots,
c_\alpha,\, k = 1, \ldots, n_\alpha \}$ of $\mathcal{H}$ such that
each family $\{ u_{a,k}^\alpha \}_{k=1}^{n_{\alpha}}$, for a given
$\alpha$, defines an orthonormal basis of $\mathcal{H}^\alpha$.
Thus, given an arbitrary orthonormal basis $\{ u_l\}_{l = 1}^n\subset\mathcal{H}$, we can compute the $n\times n$
unitary matrix $C$ with entries $C_{a,kl}^\alpha$ such that
\begin{equation}\label{CG_matrix}
u_l = \sum_{\alpha, a, k} C_{a,kl}^\alpha u_{a,k}^\alpha \, ,  \qquad  \alpha \in \widehat{H}, \quad a = 1,\ldots, c_\alpha, \quad k= 1, \ldots, n_\alpha,\quad l=1,\ldots,n \, .
\end{equation}
The coefficients $ C_{a,kl}^\alpha$ of the matrix $C$  are usually
expressed as the symbol $(l \mid \alpha, a, k)$ and are called the
\textit{Clebsch--Gordan coefficients} of the decomposition.

The original Clebsch--Gordan problem  has its origin in the
composition of two quantum systems possessing the same symmetry
group: let $\mathcal{H}_A$ and $\mathcal{H}_B$ denote Hilbert spaces
corresponding respectively to two quantum systems $A$ and $B$, which
support respective irreps $U_A$ and $U_B$ of a Lie group $G$. Then, the
composite system, whose Hilbert space is $\mathcal{H} =
\mathcal{H}_A \widehat{\otimes} \mathcal{H}_B$, supports an irrep
of the product group $G \times G$.    The interaction between both
systems gives rise to a only remaining subgroup $H \subset G\times G$ as a symmetry group of
the composite system (in  many instances, it is just $H = G$ with $G$
considered as the diagonal subgroup $G \subset
G\times G$ of the product group).    The tensor product
representation $U_A \otimes U_B$ will no longer be irreducible
with respect to the subgroup $H \subset G \times G$ and we will be
compelled to consider its decomposition into irrep components.

A considerable effort has been put in computing the
Clebsch--Gordan matrix for various situations of physical
interest. For instance, the groups $SU(N)$ have been widely
discussed (see [\onlinecite{Al10}], [\onlinecite{Gl07}] and
references therein) since when
considering the groups $SU(3)$ and $SU(2)$, the Clebsch--Gordan
matrix provides the multiplet structure and the spin components of
a composite system of particles (see [\onlinecite{Ro97}], [\onlinecite{Wi94}]). However,
all these results depend critically on the algebraic structure of
the underlying group $G$ (and the subgroup $H$) and no algorithm
was known so far to efficiently compute the
Clebsch--Gordan matrix for a general subgroup $H \subset G$ of an
arbitrary compact group $G$.

On the other hand, the problem of determining the decomposition of
an irreducible representation with respect to a given subgroup has
not been addressed from a numerical point of view.    The
multiplicity of a given irreducible representation
$(\mathcal{H}^\alpha, U^\alpha)$ of the compact group $G$ in the
finite-dimensional representation $(\mathcal{H}, U)$ is given by
the inner product
$$
c_\alpha =\frac{1}{|G|}\langle \chi^\alpha, \chi \rangle \, ,
$$
where $\chi^\alpha (g) = \mathrm{Tr}(U^\alpha(g))$ and $\chi(g) =
\mathrm{Tr} (U(g))$, $g \in G$, denote the characters of the
corresponding representations, $|G|$ is the order of the group $G$ and $\langle \cdot , \cdot \rangle$
stands for the standard inner product of central functions with
respect to the (left-invariant) Haar measure on $G$.     Hence, if
the characters $\chi^\alpha$ of the irreducible representations of
$G$ are known, the computation of the multiplicities becomes, in
principle, a simple task.  Moreover, given the characters
$\chi^\alpha$ of the irreducible representations, the projector
method would allow us to explicitly construct  the Clebsch--Gordan
matrix [\onlinecite{Tu85}, Ch.\,4].  However, if the irreducible
representations of $H$ are not known in advance (or are not
explicitly described), there is no an easy way of determining the
multiplicities $c_\alpha$.

Again, at least in principle, the computation of the  irreducible
representations of a finite group could be achieved by
constructing its character table, i.e., a $c\times c$ unitary
matrix where $c$ is the number of conjugacy classes of the group, but again, there is no a general-purpose numerical algorithm for doing that.

Recent developments in quantum group tomography  require dealing
with a broad family of representations of a large class of groups,
compact or not,  and their subgroups (see [\onlinecite{Ib09}] and
references therein for a recent overview on the subject).  Quantum
tomography allows to extend ideas from standard classical
tomography to analyze states of quantum systems.   One
implementation of quantum tomography is quantum group tomography.
Quantum group tomography is based on quantum systems supporting
representations of groups.  Such representations make it possible to
construct the corresponding \textit{tomograms} for given quantum states
[\onlinecite{Ar05}, \onlinecite{Ib11}, \onlinecite{LY15}].    Hence it is becoming
increasingly relevant to have new tools to efficiently handle
group representations and their decompositions.

It turns out that it is precisely the ideas and methods from
quantum tomography which provide the clue for the numerical
algorithm presented in this work. More explicitly, \textit{mixed quantum
states}, i.e., density matrices \textit{adapted} to a given representation,
will be used to compute the Clebsch--Gordan
matrix. \textbf{Section \ref{sec:preliminaries}} will be devoted to introduce the problem we want to solve. \textbf{Section \ref{sec:General_outline}} presents several results which will help us to show the correctness of the algorithm. The details of the numerical algorithm are contained in \textbf{Section \ref{the_algorithm:sec}}, while \textbf{Section \ref{sec:examples}} covers various examples and applications of the algorithm, among them, the decomposition of regular representations of any finite group and the decomposition of multipartite systems of spin particles.

It is remarkable that the algorithm proposed here does not
require an \textit{a priori} knowledge of the irreducible
representations of the groups and the irreducible representations
themselves are returned as outcomes of the algorithm. This makes
 the proposed algorithm an effective tool for computing the
irreducible representations, in principle, for any finite or compact group. For the sake of
clarity, most of the analysis will be done in the case of finite
groups, however it should be noted that all statements and proofs
can be easily lifted  to
compact groups by replacing finite sums over group elements by the
corresponding integrals over the group with respect to the
normalized Haar measure on it. Some additional remarks and
outcomes will be discussed at the end in
\textbf{Section \ref{sec:discussion}}. A final \hyperref[Appendix_ps]{\textbf{Appendix}} contains numerical
results for the examples addressed in \textbf{Section \ref{sec:examples}}.


\section{The setting of the problem}  \label{sec:preliminaries}


Let $G$  be a finite group of order $|G| = s$ and let $H
\subset G$  be a subgroup of $G$, not necessarily normal, of order $|H| =
r$.   We  label the elements of $G$ as  \,$G = \{ e =
h_0, g_1 = h_1, \ldots, g_{r-1}= h_{r-1}, g_r, \ldots, g_{s-1} \}
$, where the first $r$ elements correspond to the elements of the
subgroup $H$, i.e., $H = \{ e = h_0, h_1, \ldots, h_{r-1}\}$.
In what follows, a generic element in the group $G$ will be simply denoted
by $g\in G$ unless some specific indexing is required.

Let $U$ be a unitary irreducible representation of $G$ on the
finite-dimensional Hilbert space $\mathcal{H}$, $n = \dim
\mathcal{H}$, and let $e_i$, $i = 1, \ldots, n$, be any given
orthonormal basis of $\mathcal{H}$.  We denote by
\begin{equation}\label{Dg_def}
D(g)=\left[D_{ij}(g)\right]_{i,j=1}^{n}
\end{equation}
the unitary matrix
associated with $U(g)$, $g\in G$, in the chosen basis, i.e.,
\begin{equation}\label{matrix_rep_D}
D_{ij}(g)=\langle e_i,U(g) e_j\rangle\,
\end{equation}
for every $i,j=1,\ldots,n$. The restriction of the
representation $U$ to the subgroup $H$, sometimes denoted by
$U\!\!\downarrow\!\! H$ and called the {\em subduced representation} of $U$ to
$H$, will be, in general, reducible even if $U$ is irreducible.
Notice that the unitary matrix associated with $U\!\downarrow\! H(h)$,
$h \in H$, is just a submatrix of $D_{ij}(h)$ obtained by
restricting ourselves to the elements of the subgroup $H$.

A mixed state on $\mathcal{H}$, also called a \textit{density
matrix},  is a $n\times n$ normalized Hermitian positive
semidefinite matrix $\rho$, i.e.,
\begin{equation}\label{rho_conditions}
\rho = \rho^\dagger\, , \qquad \rho \geq 0 \, , \qquad \mathrm{Tr}(\rho) = 1 \,.
\end{equation}
If the unitary representation $U$ of $G$ is irreducible, then any
state $\rho$ can be written as:
\begin{equation}\label{rho_expansion}
\rho = \frac{n}{|G|} \sum_{g\in G} \mathrm{Tr}\left(\rho\, D (g)^\dagger\right) \, D(g) \,.
\end{equation}
To prove this formula one may use Schur's orthogonality relations:
\begin{equation}\label{Dg_alpha_def}
\sum_{g\in G} D^\alpha_{mn}(g)^* D^\beta_{pq}(g)= \frac{|G|}{n_\alpha} \delta_{\alpha \beta} \delta_{mp} \delta_{nq} \, ,
\end{equation}
where $*$ stands for the complex conjugate and elements $D^\alpha_{mn}(g)$ and $D^\beta_{mn}(g)$ denote, respectively,
the entries of the unitary matrices $D^{\alpha}(g)$ and $D^{\beta}(g)$ associated with the
irreducible representations $(\mathcal{H}^\alpha, U^\alpha)$ and
$(\mathcal{H}^\beta, U^\beta)$ of the group $G$ with respect to
 given arbitrary orthonormal bases in $\mathcal{H}^\alpha$ and
$\mathcal{H}^\beta$.

Let us now consider  a state $\rho$ satisfying the orthogonality
relations
\begin{equation}
\mathrm{Tr}(\rho \, D(g_k)) = 0 \, , \qquad k = r,\ldots, s-1 \, .
\end{equation}
Clearly, because of eq.\,(\ref{rho_expansion}), such a state verifies
\begin{equation}\label{adapted}
\rho = \frac{n}{|G|} \sum_{h\in H} \mathrm{Tr}\left(\rho\, D (h)^\dagger\right) \, D(h) \, .
\end{equation}
\begin{definition}\label{adapted_def}
A state $\rho$ in the Hilbert space $\mathcal{H}$ supporting an
irrep $U$ of the group $G$ is said to be \textbf{adapted} to a closed subgroup
$H$ if $\mathrm{Tr}(\rho \, D(g)) = 0$ for $g \notin H$.
\end{definition}

In other words, a state $\rho$ adapted to the subgroup $H$ of the finite
group $G$ must be of the form: 
\begin{equation*}
\rho =\frac{n}{|G|} \sum_{i=0}^{r-1} \mathrm{Tr}\left(\rho\, D (h_i)^\dagger\right) \, D(h_i) \,,
\end{equation*}
even if the subduced representation $U\!\downarrow\! H$ is reducible. 

In view of the prominent
role they will play in the algorithm, let us now
discuss briefly the role of the inner products $\mathrm{Tr}(\rho
A)$ in the realm of quantum theory: given a linear operator $A$ on $\mathcal{H}$ and a state
$\rho$, the number $\mathrm{Tr}(\rho  A)$ is called the expected
value of the operator $A$ in the state $\rho$ and it is denoted
consequently as $\langle A \rangle_\rho$.   If the operator $A$ is
self-adjoint, the expected value $\langle A \rangle_\rho$ is a
real number and it truly represents the expected value of
measuring the observable described by the operator $A$ on a
quantum system in the state $\rho$.

In the language of quantum tomography, the group function
$\chi_\rho\,\colon G \to \mathbb{C}$ defined by the
coefficients in the expansion written in eq.\,(\ref{rho_expansion}),
\begin{equation}\label{smeared_character}
\chi_\rho (g) = \mathrm{Tr}(\rho\,  D(g)) \, , \qquad g \in G \, ,
\end{equation}
is called the \textit{characteristic function} of the  state
$\rho$ associated with the representation $(\mathcal{H}, U)$ or,
depending on the emphasis, the
\textit{smeared character} of the representation $U$ with respect
to the state $\rho$ (see [\onlinecite{LY15}]). One can easily check that the
characteristic function $\chi_\rho$ is always  positive semidefinite, i.e.,
\begin{equation}\label{positivity}
\sum_{j,k= 1}^N \xi_j^* \xi_k\,  \chi_\rho (g_j^{-1}g_k) \geq 0 \, ,
\end{equation}
for all $N \in \mathbb{N}$, $\xi_j,\xi_k \in \mathbb{C}$ and $g_j,g_k \in G$.

Notice that if the state is $\rho =
\frac{1}{n}\mathds{1}$, the characteristic function
$\chi_\rho$ is the standard character $\chi (g)$ of the
representation $D(g)$ divided by $n$.  Moreover, if the representation $D(g)$ is
the trivial one, then $\chi_\rho (g) = 1$ for all $g\in G$.

\begin{definition}\label{CG_matrix_def} Let $G$ be a group, $(\mathcal{H},U)$ an irreducible
unitary representation of $G$ and $H$ a closed subgroup of $G$. The \textbf{Clebsch--Gordan matrix}
associated with $G$, $H$ and $(\mathcal{H},U)$ is the $n\times n$ matrix $C$ such that
\textnormal{\begin{equation*}
C^\dagger D(h) C=\begin{pmatrix}
                   \mathds{1}_{c_1}\otimes  D^1(h) &  &   \\
                    &  \hspace{-2cm}\mathds{1}_{c_2}\otimes  D^2(h) &  &  &  \\
                    &  & \hspace{-2.cm}\diagdots[-53]{1.6em}{.11em} &  & \hspace{0.cm}\mbox{\Huge{$0$}}&   \\
                    &  &  & \hspace{-1.3cm}\diagdots[-53]{1.6em}{.11em}  &  &  \\
                    & \hspace{-2.8cm}\mbox{\Huge{$0$}} & & & \hspace{-0.9cm}\diagdots[-53]{1.6em}{.11em}  &   \\
                    &  &  &  &  &  \hspace{-1.25cm}\mathds{1}_{c_N}\otimes  D^N(h)
\end{pmatrix},
\end{equation*}}for every $h\in H$, where the $D(h)$ are the matrices defined in $(\ref{Dg_def})$, the $D^{\alpha}(h),\ \alpha=1,\ldots,N$, are the matrices associated with the irreps of the subgroup $H$ and $\otimes$ stands for the matrix Kronecker product defined as:
\textnormal{\begin{equation*}
A\otimes B=\begin{pmatrix}
 a_{11}B & \hspace{-0.1cm}a_{12}B&\diagdots[0]{1.4em}{.12em} &\diagdots[0]{1.4em}{.12em}& a_{1n}B \\
 a_{21}B  &\hspace{-0.1cm} a_{22}B&\diagdots[0]{1.4em}{.12em} &\diagdots[0]{1.4em}{.12em}&a_{2n}B\\
\hspace{-0.2cm} \diagdots[90]{1.2em}{.12em} & \hspace{-0.5cm}\diagdots[90]{1.2em}{.12em}&\hspace{-1.2cm}\diagdots[-42]{1.6em}{.11em}& &\hspace{-0.2cm}\diagdots[90]{1.2em}{.12em}   \\
\hspace{-0.2cm} \diagdots[90]{1.2em}{.12em} & \hspace{-0.5cm} \diagdots[90]{1.2em}{.12em}&\hspace{0.41cm}\diagdots[-42]{1.6em}{.11em}&   &\hspace{-0.2cm}\diagdots[90]{1.2em}{.12em}\\
 \vspace{0.1cm}\hspace{-0.2cm}\diagdots[90]{0.85em}{.1em} & \hspace{-0.5cm} \diagdots[90]{0.85em}{.1em} & &\hspace{0.52cm}\diagdots[-42]{1.6em}{.11em}& \hspace{-0.2cm}\diagdots[90]{0.85em}{.10em}\\
   a_{m1}B&  \hspace{-0.1cm}a_{m2}B&\diagdots[0]{1.4em}{.12em} &\diagdots[0]{1.4em}{.12em} &a_{mn}B
\end{pmatrix}
\end{equation*}}for arbitrary matrices $A=\left(a_{ij}\right)_{i,j=1}^{m,n}$ and $B$.
\end{definition}

Since the unitary representation is unique (modulo unitary transformations within each proper invariant subspace ${\mathcal H}^{ \alpha}$ or
permutations among the  ${\mathcal H}^{ \alpha}$), the Clebsch--Gordan matrix is also unique (except for such transformations), (see [\onlinecite{Tu85}] for more detailed information about this).

Finally, let us specify the kind of adapted states we will be using in the algorithm. As we shall see,
such states will have to satisfy certain nondegeneracy conditions.  

Given any adapted state $\rho$,
we know that, according to (\ref{adapted}), $\rho$ is a linear combination of the representations $D(h),\ h\in H$, therefore the Clebsch--Gordan matrix $C$ in \textbf{Def.\,\ref{CG_matrix_def}} will block-diagonalize $\rho$
in the form:
\begin{equation}\label{structure_rho}
C^\dagger \rho\,C=\begin{pmatrix}
                   \mathds{1}_{c_1}\otimes  \sigma^1 &  &   \\
                    &  \hspace{-1.2cm}\mathds{1}_{c_2}\otimes  \sigma^2 &  &  &  \\
                    &  & \hspace{-0.6cm}\diagdots[-53]{1.6em}{.11em} &  & \hspace{0.4cm}\mbox{\Huge{0}}&   \\
                    &  &  & \hspace{-0.cm}\diagdots[-53]{1.6em}{.11em}  &  &  \\
                    & \hspace{-0.8cm}\mbox{\Huge{0}} & & & \hspace{-0.1cm}\diagdots[-53]{1.6em}{.11em}  &   \\
                    &  &  &  &  &  \hspace{-1.cm}\mathds{1}_{c_N}\otimes  \sigma^N
\end{pmatrix},
\end{equation}
where each block $\sigma^{\alpha},\ \alpha=1,\ldots,N$, is a Hermitian positive semidefinite matrix of the same dimension as
the corresponding $D^{\alpha}(h)$.
Now, consider  the spectral decomposition of the matrices $\sigma^\alpha$, i.e.,
\begin{equation}\label{eigen}
\sigma^\alpha r_j^\alpha=\lambda_j^\alpha r_j^\alpha,\qquad \langle r^\alpha_j,r^\alpha_k\rangle=\delta_{jk}\,, \quad j,k=1,\ldots,n_\alpha,
\end{equation}
where the $r_j^\alpha$ are orthonormal eigenvectors of $\sigma^\alpha$  within  each proper
subspace ${\mathcal H}^{\alpha}, \ \alpha=1,\ldots,N.$

\begin{definition}\label{generic_def}
An adapted state $\rho$ is said to be \textbf{generic} if its eigenvalues  have the minimum possible
degeneracy, that is, $\lambda_j^\alpha\neq\lambda_k^\beta$ for all
$\alpha,\beta=1,\ldots,N$, and for all $j=1,\ldots,n_\alpha$,
$k=1,\ldots,n_\beta$.
\end{definition}
Notice that the eigenvalues cannot in general be simple since each $\lambda_j^\alpha$ has by
construction multiplicity $c_{\alpha}$ (recall eq.\,(\ref{structure_rho})). In the contruction of the algorithm, a further concept of pair-wise genericity will be needed:

\begin{definition}\label{generic_pair_def}
A pair $(\rho_1,\rho_2)$ of  adapted states is said to be \textbf{mutually generic} if they are both generic (in the sense of {\bf{Definition \ref{generic_def}}}) and no eigenvector $r^\alpha_{j}$ of the block $\sigma_1^\alpha$ of $\rho_1$ is an eigenvector of the corresponding
$\sigma_2^\alpha$ of $\rho_2$ whenever $n_{\alpha}>1$, where matrices $\sigma^\alpha_a$ come from the block-diagonalization of the adapted states $\rho_a$: 
\textnormal{$$
C^\dagger \rho_a C=\diag(\mathds{1}_{c_1}\otimes  \sigma_a^1,\mathds{1}_{c_2}\otimes  \sigma_a^2,\ldots,\mathds{1}_{c_N}\otimes  \sigma_a^N),\quad a=1,2.
$$}
\end{definition}

Of course, we exclude the case $n_\alpha=1$ in which the proper invariant subspace has dimension one and therefore, the eigenvectors must coincide.


\section{General outline}\label{sec:General_outline}

Before we provide a detailed description of the decomposition
algorithm we  propose, let us first give a rough outline of how
the algorithm is organized and, especially, why it works.

The final goal of the algorithm is to find the Clebsch--Gordan matrix $C$,
which, as shown in \textbf{Def.\,\ref{CG_matrix_def}}, block-diagonalizes all the elements of the representation $D(h)$, $h\in H$. In other words, the columns of $C$ provide orthonormal bases for all proper invariant subspaces \,${\mathcal H}^{\alpha}$, which are common to all $D(h)$, $h\in H$ (and consequently, common to all adapted states).

Now, consider any fixed adapted state $\rho$ and any unitary matrix $V$ diagonalizing $\rho$ pointwise,
i.e., such that $V^{\dagger}\rho V$ is diagonal. The idea underlying our algorithm is that since
the columns of both $V$ and $C$ span the same proper invariant subspaces, they must be somehow related.
This connection, which is crucial to our argument, will be made explicitly in
\textbf{Theorem~\ref{triple_V}} below, and implies that, after appropriate reordering of the columns of $V$,
any other adapted state (more generally, any matrix which is a linear combination of the $D(h)$, $h\in H$) will be {\em block}-diagonalized by $V$ (see \textbf{Corollary~\ref{Transformation_any_state}} below). Furthermore, the diagonal blocks one obtains have a very particular structure which, once identified in \textbf{Corollary~\ref{Transformation_any_state}}, will be the key to extract the Clebsch--Gordan matrix $C$ out of $V$ via appropriate similarity transformations described both in \textbf{Corollary~\ref{R_tilde:corollary}} and \textbf{Lemma~\ref{permutation:lemma}}.

The following result is the foundation of the algorithm we describe in \textbf{Section \ref{the_algorithm:sec}} below:

\begin{theorem}\label{triple_V} Let $\rho$ be any generic adapted state and let $V$ be any unitary matrix
such that $V^{\dagger}\rho V$ is diagonal. Then,
$$
V=CXP,
$$
where $C$ is the Clebsch--Gordan matrix defined as in {\bf Definition \ref{CG_matrix_def}}, $P$ is any permutation matrix and
$X=\diag(X^1,X^2,\ldots,X^N)$ with $X^\alpha$ given by
\textnormal{\begin{equation}\label{def_X}
X^\alpha=\left(\begin{array}{c|c|c|c}
Q^\alpha_1\otimes r^{\alpha}_1&Q^\alpha_2\otimes r^{\alpha}_2&\cdots&Q^\alpha_{n_\alpha}\otimes r^{\alpha}_{n_\alpha}
\end{array}\right),
\end{equation}}for any set of $c_\alpha\times c_\alpha$ unitary matrices
$\left\{Q^\alpha_j\right\}_{j=1}^{n_\alpha}$\,, where
 $\left\{r^\alpha_{j}\right\}_{j=1}^{n_\alpha}$ is a set
of eigenvectors of the matrices $\sigma^\alpha$,
$\alpha=1,\ldots,N$, given in eq.\,$(\ref{eigen})$.
\end{theorem}

\noindent\textbf{Proof}: It follows from (\ref{eigen}) that
\begin{equation*}
(\mathds{1}_{c_\alpha}\otimes \sigma^{\alpha})(z_j^p\otimes r^{\alpha}_j)=\lambda^\alpha_{j}z_j^p\otimes r^{\alpha}_j\,
\end{equation*}
for any choice of $n_\alpha$ orthonormal bases $\left\{z_j^p\right\}_{p=1}^{c_\alpha}$\,, $j=1,\ldots,n_\alpha$.
Recall that $n_\alpha$ is the dimension of the invariant subspace $\mathcal{H}^\alpha$ or,
equivalently, the number of rows  and columns of the Hermitian positive semidefinite  matrices $\sigma^\alpha$.
On the other hand, $c_\alpha$ is the multiplicity of that subspace, i.e.,  the global multiplicity of
the eigenvalues $\lambda^\alpha_j$ in the total matrix $\rho$ (see eq.\,(\ref{structure_rho})).

If we now construct unitary matrices:
\begin{equation*}
Q^\alpha_j=\begin{pmatrix}
|&| & &|\\
z^1_j&z^2_j&\cdots&z^{c_\alpha}_j\\
|&| & &|
\end{pmatrix},
\end{equation*}
such that their columns  are
the orthonormal vectors of the basis
$\displaystyle{\left\{z_j^p\right\}_{p=1}^{c_\alpha}}$, then
the matrix
\begin{equation}\label{X_alpha_def}
X^\alpha=\left(\begin{array}{c|c|c|c}
\!\!Q^\alpha_1\otimes r^{\alpha}_1&Q^\alpha_2\otimes r^{\alpha}_2&\cdots&Q^\alpha_{n_\alpha}\otimes r^{\alpha}_{n_\alpha}\!\!\!
\end{array}\right)
\end{equation}
will diagonalize the matrix $\mathds{1}_{c_\alpha}\otimes\sigma^\alpha$ with its eigenvalues sorted as follows:
\begin{equation}\label{lambda_alpha_prove}
{X^\alpha}^\dagger(\mathds{1}_{c_\alpha}\otimes\sigma^{\alpha})X^\alpha=\begin{pmatrix}
                   \lambda_{1}^\alpha\mathds{1}_{c_\alpha} &  &   \\
                    &  \hspace{-0.8cm}\lambda_{2}^\alpha\mathds{1}_{c_\alpha} &  &  &  \\
                    &  & \hspace{-0.6cm}\diagdots[-47]{1.4em}{.12em} &  & \hspace{0.2cm}\mbox{\Huge{0}}&   \\
                    &  &  & \hspace{0.1cm}\diagdots[-47]{1.4em}{.12em} &  &  \\
                    & \mbox{\Huge{0}} & & & \hspace{0.45cm}\diagdots[-47]{1.4em}{.12em} &   \\
                    &  &  &  &  &  \hspace{-0.4cm}\lambda_{n_\alpha}^\alpha\mathds{1}_{c_\alpha}
\end{pmatrix}=\Lambda^\alpha\,.
\end{equation}
Therefore, in view of (\ref{structure_rho}), the matrix $X=\diag(X^1,X^2,\ldots,X^N)$ diagonalizes the matrix $C^\dagger\rho C$:
\begin{equation*}
(CX)^\dagger\rho\,CX=\diag\left(\Lambda^1,\Lambda^2,\ldots,\Lambda^N\right),
\end{equation*}
and any permutation $P$ of the columns of the matrix $CX$
will still diagonalize $\rho$, which shows that any unitary matrix $V$
diagonalizing $\rho$ can be written as a product $V=CXP$.  

\hfill $\Box$

\begin{corollary}\label{Transformation_any_state}
Let $\rho$ be any adapted state, let $X$ be the associated block-diagonal matrix with blocks $(\ref{def_X})$,
let $P=\diag(P^{1},P^2,\ldots,P^{N})$ with $P^{\alpha}=\diag(P^\alpha_1,P^\alpha_2,\ldots,P^\alpha_{n_\alpha}),\ \alpha\in\{1,\ldots,N\}$, where each $P^{\alpha}$ is
a $c_{\alpha}n_{\alpha}\times c_{\alpha}n_{\alpha}$ permutation matrix and let $V=CXP$.
Then, for any linear combination $\tau={\displaystyle \sum_{h\in H}\alpha_h D(h)}$, it is verified that
\textnormal{\begin{equation*}\label{structure_tau}
\hspace{0cm}V^\dagger \tau V=\left(
\begin{array}{ccccc}
\left.\begin{array}{|cccc|}\hline
 \phantom{a} &  \phantom{a}&  \phantom{a}&  \phantom{a}\\
 \phantom{a} &  \phantom{a}&  \phantom{a}&  \phantom{a}\\\hline
\end{array}\right\}c_1n_1
 & &  &  &\\
 &\hspace{-1.33cm} \left.\begin{array}{|ccccccc|}\hline
 \phantom{a}& \phantom{a} & \phantom{a} &  \phantom{a}&  \phantom{a}&  \phantom{a}&  \phantom{a}\\
  \phantom{a}&  \phantom{a}&  \phantom{a}& \phantom{a}&  \phantom{a}&  \phantom{a}&  \phantom{a} \\
  \phantom{a}& \phantom{a} & \phantom{a} & \phantom{a}&  \phantom{a}&  \phantom{a}&  \phantom{a} \\
  \phantom{a}& \phantom{a} & \phantom{a} & \phantom{a}&  \phantom{a}&  \phantom{a}&  \phantom{a}\\\hline
\end{array}\right\} c_2n_2 &   & &\\
& &\hspace{-1.8cm}\diagdots[-46]{1.6em}{.11em} & &\\
& & &\hspace{-0.73cm}\diagdots[-46]{1.6em}{.11em} &\\
& & & &\hspace{-1.6cm}c_Nn_N\left\{\begin{array}{|ccccc|}\hline
 \phantom{a} &  \phantom{a}& \phantom{a}&  \phantom{a}&  \phantom{a}\\
 \phantom{a} &  \phantom{a}& \phantom{a}&  \phantom{a}&  \phantom{a}\\
 \phantom{a} &  \phantom{a}& \phantom{a}&  \phantom{a}&  \phantom{a}\\\hline
\end{array}\right.
 \end{array}
\right),\begin{array}{c}
\hspace{-8.35cm}\vspace{6.8cm}\Sigma^1
\end{array}
\phantom{a}
\begin{array}{c}
\hspace{-6.2cm}\vspace{2.3cm}\Sigma^2
\end{array}
\phantom{a}
\begin{array}{c}
\hspace{-2.45cm}\vspace{-5.9cm}\Sigma^N
\end{array}
\end{equation*}}where
\textnormal{\begin{equation*}\label{Sigma_structure_alg}
\Sigma^\alpha=\begin{pmatrix}
 R_{11}^\alpha & \hspace{-0.2cm}R_{12}^\alpha&\diagdots[0]{1.4em}{.12em} &\diagdots[0]{1.4em}{.12em}& R_{1n_\alpha}^\alpha \\
 R_{21}^\alpha  &\hspace{-0.2cm} R_{22}^\alpha&\diagdots[0]{1.4em}{.12em} &\diagdots[0]{1.4em}{.12em}&R_{2n_\alpha}^\alpha\\
\hspace{0.1cm} \diagdots[90]{1.2em}{.12em} & \hspace{-0.2cm}\diagdots[90]{1.2em}{.12em}&\hspace{-0.9cm}\diagdots[-43]{1.6em}{.11em}& &\hspace{-0.15cm}\diagdots[90]{1.2em}{.12em}   \\
\hspace{0.1cm} \diagdots[90]{1.2em}{.12em} & \hspace{-0.2cm} \diagdots[90]{1.2em}{.12em}&\hspace{0.65cm}\diagdots[-43]{1.6em}{.11em}&   &\hspace{-0.15cm}\diagdots[90]{1.2em}{.12em}\\
 \vspace{0.1cm}\hspace{0.1cm}\diagdots[90]{0.85em}{.1em} & \hspace{-0.2cm} \diagdots[90]{0.85em}{.1em} & &\hspace{0.47cm}\diagdots[-43]{1.6em}{.11em}& \hspace{-0.15cm}\diagdots[90]{0.85em}{.10em}\\
   R_{n_\alpha 1}^\alpha&  \hspace{-0.2cm}R_{n_\alpha 2}^\alpha&\diagdots[0]{1.4em}{.12em} &\diagdots[0]{1.4em}{.12em} &R_{n_\alpha n_\alpha}^\alpha
\end{pmatrix},
\end{equation*}}with $R_{ij}^\alpha$ square matrices of size $c_\alpha$ defined as:
\textnormal{\begin{equation*}\label{Rij_structure}
R_{ij}^\alpha=s_{ij}^\alpha {(Q_i^\alpha P_i^\alpha)}^\dagger Q_j^\alpha P_j^\alpha\qquad\mbox{for}\qquad s_{ij}^\alpha={r^{\alpha}_i}^\dagger\tau^{\alpha} r^{\alpha}_j\,,
\end{equation*}}where $\tau^\alpha$, $\alpha=1,\ldots,N$, are the matrices
on the block diagonal of $\tau$ after being transformed
by $C$, i.e., those matrices such that
$C^\dagger\tau
C=\diag\big(\mathds{1}_{c_1}\otimes\tau^1,\mathds{1}_{c_2}\otimes\tau^2,\ldots,\mathds{1}_{c_N}\otimes\tau^N\big)$\,.
\end{corollary}

\noindent\textbf{Proof}: We just transform $\tau$ with $V$:
\begin{equation*}
V^\dagger\tau V=\begin{pmatrix}
                   \big(X^1 P^1\big)^\dagger(\mathds{1}_{c_1}\otimes\tau^1)X^1 P^1&  &   \\
                    &  \hspace{-2.4cm}\big(X^2 P^2\big)^\dagger(\mathds{1}_{c_2}\otimes\tau^2)X^2 P^2&  &  &  \\
                    &  & \hspace{-1.85cm}\diagdots[-33]{1.6em}{.12em} &  & \hspace{2cm}\mbox{\Huge{0}}&   \\
                    &  &  & \hspace{-0.cm}\diagdots[-33]{1.6em}{.12em}  &  &  \\
                    & \hspace{-2.8cm}\mbox{\Huge{0}} & & & \hspace{-0.27cm}\diagdots[-33]{1.6em}{.12em}  &   \\
                    &  &  &  &  &  \hspace{-3.5cm}\big(X^NP^N\big)^\dagger(\mathds{1}_{c_N}\otimes\tau^N)X^NP^N
\end{pmatrix}.
\end{equation*}
Hence, the matrices $\Sigma^{\alpha}$ in the statement are \,$\displaystyle{\Sigma^\alpha=\big(X^\alpha P^{\alpha}\big)^\dagger(\mathds{1}_{c_1}\otimes\tau^\alpha)X^\alpha P^{\alpha}}$. Finally, if we substitute in $\displaystyle{\Sigma^\alpha}$ the definition of $\displaystyle{X^\alpha}$ in eq.\,(\ref{X_alpha_def}) and use the property  $(A\otimes B)(C\otimes D)=AC\otimes BD$ of the Kronecker product for matrices $A,B,C,D$ such that the products $AC$ and $BD$ are feasible , we get:
$$
\displaystyle{R_{ij}^\alpha=s_{ij}^\alpha {P_i^\alpha}^\dagger{Q_i^\alpha}^\dagger Q_j^\alpha P_j^\alpha} \qquad \mbox{\rm with} \qquad  \displaystyle{s_{ij}^\alpha={r^{\alpha}_i}^\dagger\tau^{\alpha} r^{\alpha}_j}.
$$
\hfill $\Box$

This corollary is key  to the algorithm described in \textbf{Section \ref{the_algorithm:sec}}
below because it means that any matrix  diagonalizing one generic adapted
state $\rho$, with the eigenvectors appropriately reordered, will
transform any linear combination of the representation $D(h)$ (in particular, any
other adapted state) into the specific form given by \textbf{Corollary \ref{Transformation_any_state}}, which has a very special structure. Our next step amounts to exploiting this
structure in order to reveal a finer block structure within each $\Sigma^{\alpha}$
for any linear combination of the representation.

\begin{corollary}\label{R_tilde:corollary}
Let $\rho,\tau, V$ and $\Sigma^{\alpha},\ \alpha\in\{1,\ldots,N\}$, be as in {\bf Corollary \ref{Transformation_any_state}}.
Let
\textnormal{$$
\widetilde{R}^{\alpha}_{ij}=\frac{R^{\alpha}_{ij}}{\|R^{\alpha}_{ij}\|}\,
$$}for any matrix $R_{ij}^\alpha\not=0$ and set
\textnormal{$$
\widetilde{R}^\alpha_{k_\alpha}=\diag\left(\widetilde{R}^{\alpha}_{1{k_\alpha}},\widetilde{R}^{\alpha}_{2{k_\alpha}},\ldots,\widetilde{R}^{\alpha}_{n_{\alpha}{k_\alpha}}\right)
$$}for any fixed $k_\alpha\in\{1,\ldots,n_{\alpha}\}$. If $\,\Xi^{\alpha},\ \alpha\in\{1,\ldots,N\}$, are the diagonal blocks
of \textnormal{$V^{\dagger} \kappa V$} for some other ${\displaystyle \kappa=\sum_{h\in H}\beta_h D(h)}$, then:
\textnormal{
\begin{equation*}
 \widetilde{R}^\alpha_{k_\alpha}\hspace{-0.44cm}\phantom{A}^\dagger\,\Xi^\alpha \widetilde{R}^\alpha_{k_\alpha}=\widetilde{S}_{k_\alpha}^\alpha\otimes\mathds{1}_{c_\alpha}=
 \begin{pmatrix}
  \tilde{s}_{{k_\alpha}11}^\alpha \mathds{1}_{c_\alpha} & \hspace{-0.2cm} \tilde{s}_{{k_\alpha}12}^\alpha \mathds{1}_{c_\alpha}&\diagdots[0]{1.4em}{.12em} &\diagdots[0]{1.4em}{.12em}&  \tilde{s}_{{k_\alpha}1n_\alpha}^\alpha \mathds{1}_{c_\alpha} \\
  \tilde{s}_{{k_\alpha}21}^\alpha \mathds{1}_{c_\alpha}  &\hspace{-0.2cm}  \tilde{s}_{{k_\alpha}22}^\alpha \mathds{1}_{c_\alpha}&\diagdots[0]{1.4em}{.12em} &\diagdots[0]{1.4em}{.12em}& \tilde{s}_{{k_\alpha}2n_\alpha}^\alpha \mathds{1}_{c_\alpha}\\
\hspace{-0.1cm} \diagdots[90]{1.2em}{.12em} & \hspace{-0.5cm}\diagdots[90]{1.2em}{.12em}&\hspace{-1.4cm}\diagdots[-30]{1.6em}{.12em}& &\hspace{-0.4cm}\diagdots[90]{1.2em}{.12em}   \\
\hspace{-0.1cm} \diagdots[90]{1.2em}{.12em} & \hspace{-0.5cm} \diagdots[90]{1.2em}{.12em}&\hspace{1.1cm}\diagdots[-30]{1.6em}{.12em}&   &\hspace{-0.4cm}\diagdots[90]{1.2em}{.12em}\\
 \hspace{-0.1cm}\diagdots[90]{1.2em}{.12em} & \hspace{-0.5cm} \diagdots[90]{1.2em}{.12em} & &\hspace{1cm}\diagdots[-30]{1.6em}{.12em}& \hspace{-0.4cm}\diagdots[90]{1.2em}{.12em}\\
    \tilde{s}_{{k_\alpha}n_\alpha 1}^\alpha \mathds{1}_{c_\alpha}&  \hspace{-0.2cm} \tilde{s}_{{k_\alpha}n_\alpha 2}^\alpha \mathds{1}_{c_\alpha}&\diagdots[0]{1.4em}{.12em} &\diagdots[0]{1.4em}{.12em} & \tilde{s}_{{k_\alpha}n_\alpha n_\alpha}^\alpha \mathds{1}_{c_\alpha}
\end{pmatrix}.
\end{equation*}}
\end{corollary}

\noindent\textbf{Proof}: If we write
$$
\Xi^\alpha =\begin{pmatrix}
 T_{11}^\alpha & \hspace{-0.3cm}T_{12}^\alpha&\diagdots[0]{1.4em}{.12em} &\diagdots[0]{1.4em}{.12em}& T_{1n_\alpha}^\alpha \\
 T_{21}^\alpha  &\hspace{-0.2cm} T_{22}^\alpha&\diagdots[0]{1.4em}{.12em} &\diagdots[0]{1.4em}{.12em}&T_{2n_\alpha}^\alpha\\
\hspace{-0.1cm} \diagdots[90]{1.2em}{.12em} & \hspace{-0.5cm}\diagdots[90]{1.2em}{.12em}&\hspace{-0.8cm}\diagdots[-40]{1.6em}{.11em}& &\hspace{-0.4cm}\diagdots[90]{1.2em}{.12em}   \\
\hspace{-0.1cm} \diagdots[90]{1.2em}{.12em} & \hspace{-0.5cm} \diagdots[90]{1.2em}{.12em}&\hspace{0.85cm}\diagdots[-40]{1.6em}{.11em}&   &\hspace{-0.4cm}\diagdots[90]{1.2em}{.12em}\\
 \vspace{0.1cm}\hspace{-0.2cm}\diagdots[90]{0.85em}{.1em} & \hspace{-0.5cm} \diagdots[90]{0.85em}{.1em} & &\hspace{0.55cm}\diagdots[-40]{1.6em}{.11em}& \hspace{-0.4cm}\diagdots[90]{0.85em}{.10em}\\
   T_{n_\alpha 1}^\alpha&  \hspace{-0.2cm}T_{n_\alpha 2}^\alpha&\diagdots[0]{1.4em}{.12em} &\diagdots[0]{1.4em}{.12em} &T_{n_\alpha n_\alpha}^\alpha
\end{pmatrix},
$$
where $T_{ij}^{\alpha}=t_{ij}^{\alpha}{(Q_i^{\alpha}P_i^\alpha)}^{\dagger}Q_j^{\alpha}P_j^\alpha$, then one can easily check that
$$
\widetilde{R}_{i{k_\alpha}}^{\alpha \dagger}T_{ij}^{\alpha}\widetilde{R}_{j{k_\alpha}}^{\alpha}= \frac{\overline{s_{i{k_\alpha}}^{\alpha}}}{|s_{i{k_\alpha}}^{\alpha}|}\,t_{ij}^{\alpha}\,\frac{s_{j{k_\alpha}}^{\alpha}}{|s_{j{k_\alpha}}^{\alpha}|}\mathds{1}_{c_{\alpha}}=\tilde{s}_{{k_\alpha}ij}^{\alpha}\mathds{1}_{c_{\alpha}}\quad\mbox{\rm and}\quad\widetilde{S}_{k_\alpha}^\alpha=(\tilde{s}_{{k_\alpha}ij}^{\alpha})_{i,j=1}^{n_\alpha}.
$$
\hfill $\Box$

\bigskip

Notice that this transformation leads to a matrix with almost the structure of (\ref{structure_rho}), with the difference that the entries in the blocks $\sigma^\alpha$ are scattered everywhere instead of being concentrated
in the diagonal blocks. In other words, if we set
\begin{equation}\label{def_r_tilde}
\widetilde{R}=\diag\left(\widetilde{R}_{k_1}^1,\widetilde{R}_{k_2}^2,\ldots,\widetilde{R}_{k_N}^{N}\right)
\end{equation}
for
$k_{\alpha}\in\{1,\ldots,n_{\alpha}\}$ such that $\widetilde{R}_{jk_\alpha}^{\alpha}\not=0$ for all
$j\in\{1,\ldots,n_{\alpha}\}$, then

\begin{equation}
\big(V\widetilde{R}\big)^\dagger\kappa V\widetilde{R}=\diag\left(\widetilde{S}_{k_1}^1\otimes\mathds{1}_{c_1},\widetilde{S}_{k_2}^2\otimes\mathds{1}_{c_2},\ldots,\widetilde{S}_{k_N}^N\otimes\mathds{1}_{c_N}\right),
\end{equation}
while we would like to have the Kronecker products in reverse order. It is well-known
that for any pair of matrices $A$ and $B$ of arbitrary dimensions, the two Kronecker
products $A \otimes B$ and $B\otimes A$ are permutationally equivalent (i.e., $B\otimes A=P( A\otimes B) F$ for
appropriate permutation matrices $P$ and $F$). Moreover, when both
$A$ and $B$ are square, they are actually permutationally similar (i.e., one can take $P=F ^{\dagger}$
above: see, for instance, \textbf{Corollary 4.3.10} in [\onlinecite{Ho91}] or [\onlinecite{He81}]).

\begin{lemma}\label{permutation:lemma}
Given two matrices $A$ and $B$ of arbitrary sizes, there exist two permutation
matrices $P$ and $F$, which only depend on the dimensions of the
matrices $A$ and $B$, such that
$$
B\otimes A=P(A\otimes B)F\,.
$$
In the case in which $A$ and $B$ are square matrices of sizes
$n$ and $c$ respectively, the permutation matrices are related by \textnormal{$P=F^\dagger$}
where
$$
F=\left(\begin{array}{c|c|c|c|c}
f &h f & h^2f &\cdots&h^{c-1}f\end{array}\right),
$$
and $h$ and $f$ are the following matrices of dimensions $cn\times cn$ and $cn\times n$ respectively:
$$
h=\begin{pmatrix}
                   0 &\hspace{-0.2cm}1  &   \\
                     & \hspace{-0.2cm}0  & \hspace{0.1cm}\diagdots[-50]{1.5em}{.11em}&  &  \\
                    &  & \hspace{0.1cm}\diagdots[-50]{1.5em}{.11em} & \hspace{0.3cm}\diagdots[-50]{1.5em}{.11em} & \hspace{0.8cm}\mbox{\Huge{$0$}}&   \\
                    &  &  &  \hspace{0.25cm}\diagdots[-50]{1.5em}{.11em} & \hspace{-0.5cm}\diagdots[-50]{1.5em}{.11em} &  \\
                    & \mbox{\Huge{$0$}} & &  &\hspace{-0.4cm}\diagdots[-50]{1.5em}{.11em} &\hspace{-0.8cm}1  \\
                    &  &  &  & & \hspace{-0.8cm}0
\end{pmatrix}\,,\quad\qquad f=\left(\begin{array}{c|c}
1&\begin{array}{c}
\vspace{-0.3cm}\\
\Large{\text{$0$}}_{1\times(n-1)}\\
\vspace{-0.4cm}
\end{array}\\\hline
\begin{array}{c}
\vspace{0.2cm}\\
\Large{\text{$0$}}_{(cn-1)\times 1}
\end{array}&\hspace{-0.12cm}\begin{array}{c}
\vspace{-0.3cm}\\
\hspace{0.2cm}\mathds{1}_{(n-1)}\otimes\left(\begin{array}{c}
0\\
\vdots\\
0\\
1
\end{array}\right)_{c\times 1}
\vspace{0.1cm}\\\hline
\end{array}\\
&\begin{array}{c}
\vspace{-0.3cm}\\
\huge{\text{$0$}}_{(c-1)\times (n-1)}
\end{array}
\end{array}\right)\,.
$$
\end{lemma}

As a consequence of \textbf{Lemma \ref{permutation:lemma}}, if we compute the matrix $\widetilde{F}=\diag\left(F^1,F^2,\ldots,F^N\right)$ such that
\begin{equation}
{F^\alpha}^\dagger \left(\widetilde{S}_{k_\alpha}^\alpha\otimes \mathds{1}_{c_\alpha}\right) F^\alpha=\left(\mathds{1}_{c_\alpha}\otimes \widetilde{S}_{k_\alpha}^\alpha\right),
\end{equation}
if $V$ is the unitary matrix in \textbf{Corollary \ref{Transformation_any_state}} and $\widetilde{R}$ is given by (\ref{def_r_tilde}), we conclude that
\begin{equation}
C=V\widetilde{R}\widetilde{F}
\end{equation}
is the Clebsch--Gordan matrix in \textbf{Definition \ref{CG_matrix_def}}.

\section{The algorithm}\label{the_algorithm:sec}

We are now in the position to give a detailed description, step by step, of the
decomposition algorithm. We first specify the input and the output of the algorithm:

\begin{itemize}
\medskip

\item \textbf{Input}: A unitary  representation
of any finite group or compact Lie group $H$.
\medskip

\item \textbf{Output}: The  Clebsch--Gordan matrix $\widehat{C}$, in a basis of eigenvectors of an initial adapted state $\rho_1$.

\end{itemize}

\medskip

We may organize the algorithm into eight steps:

\begin{enumerate}

\phantomsection\label{step_1}\item[1.]  {\bf Generate two  adapted states}: We start by creating two mutually generic states $\rho_1$ and $\rho_2$ (see \textbf{Definition \ref{generic_pair_def}}).
To create them, we generate  two random vectors $\boldsymbol{\varphi}_1$ and $\boldsymbol{\varphi}_2$ of size $r=|H|$ with no zero components and use their
respective entries as coefficients to construct two linear combination of the matrices $D(h),\ h\in H$:
\begin{equation*}\label{rho_tilde}
\tau_{a}=\sum_{j=0}^{r-1}\varphi_{a}(j)D(h_j)\,,\qquad a=1,2\,.
\end{equation*}
Next, we symmetrize:
\begin{equation*}
\tilde{\rho}_{a}=\tau_{a}+\tau_{a}^\dagger\,,
\end{equation*}
shift them by the spectral radius and divide by the trace:
\begin{equation*}\label{adapted_state}
\tilde{\rho}'_{a}=\tilde{\rho}_{a}+s_{\textrm{radius}}(\tilde{\rho}_{a})\mathds{1},\qquad \rho_{a}=\frac{\tilde{\rho}'_{a}}{\mathrm{Tr}(\tilde{\rho}'_{a})}\,,\qquad a=1,2,
\end{equation*}
to obtain two Hermitian normalized positive semidefinite matrices $\rho_1$ and $\rho_2$.
Having been randomly generated, it is safe to assume that they are mutually generic.

\bigskip
\item[2.]  \textbf {Diagonalize pointwise the first state}: Compute a unitary matrix $V_1$ which diagonalizes pointwise  the state $\rho_1$, i.e.,
such that $V_1^{\dagger}\,\rho_1\,V_1$ is a diagonal
matrix. Such matrix exists since $\rho_1$ is Hermitian.

\bigskip
\phantomsection\label{step_3}\item[3.]  \textbf {First sorting}: Reorder the columns of $V_1$ by grouping together the eigenvectors
corresponding to the same proper subspace $\mathcal{L}^\alpha$. Recall that, according to \textbf{Corollary \ref{Transformation_any_state}},
there is a reordering of the columns of $V_1$ which  block-diagonalizes $\rho_2$ and the dimensions of the diagonal blocks are the
dimensions of the $\mathcal{L}^\alpha$. Notice that,  if two columns $v_j$ and $v_k$ of $V_1$ correspond to the same proper subspace $\mathcal{L}^\alpha$,
then $v_j^\dagger\rho_2 v_k\neq 0$. This will be our test for rearranging the columns of $V_1$. More precisely, we  use  the
following routine, based on a divide-and-conquer approach:

\begin{enumerate}
\medskip
\phantomsection\label{step_3_1}\item[3.1.] Choose one column of $V_1$, rename it as $v_1^{sort}$ and move it into a list of vectors we will call $L^{sort}$.
\medskip

\begin{figure}[htbp]
\includegraphics{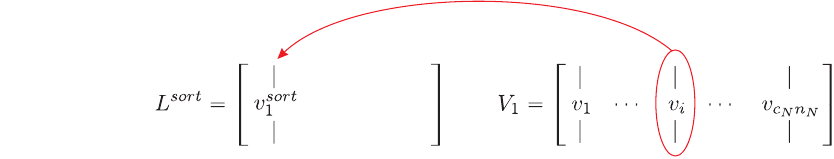}
\caption{\hspace{2.3cm}{STEP 3.1}. Choosing the starting vector.}
\end{figure}

\phantomsection\label{step_3_2}\item[3.2.] Compute $\epsilon_{1k}={v_1^{sort}}^\dagger\rho_2 v_k$ for another column $v_k$ of $V_1$ and if $\epsilon_{1k}\neq 0$, move $v_k$ into the list  $L^{sort}$ and rename it as $v_2^{sort}$.
Repeat on all remaining columns of $V_1$, move those $v_k$ with ${v_1^{sort}}^\dagger\rho_2 v_k\not=0$ into the list $L^{sort}$ and label them as $v_j^{sort}$, with the index $j$ reflecting
the order in which they have been included into the list.
\begin{figure}[htbp]
\includegraphics{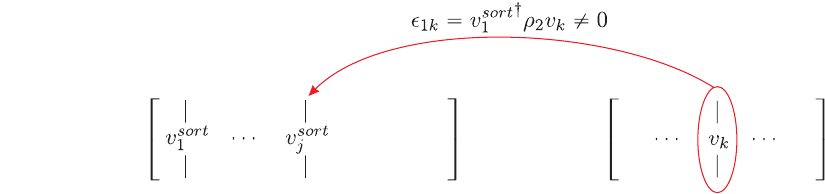}
\caption{\hspace{2.4cm}{STEP 3.2}. Finding vectors in the same subspace as $v_1^{sort}$.}
\end{figure}

\item[3.3.] Compute $\epsilon_{jk}={v_j^{sort}}^\dagger\rho_2 v_{k}$ for $v_j^{sort}\in L^{sort},\ j\geq 2$, for those
columns $v_k$ of $V_1$ not yet moved into $L^{sort}$ in STEP \hyperref[step_3_2]{3.2}.
This is a re-check since there might be some vector left not included in the list in STEP \hyperref[step_3_2]{3.2} because
it happened to be orthogonal to $v_1^{sort}$ in the scalar product defined by $\rho_2$.
The mutual genericity condition ensures that no vector in $L^{sort}$ can be orthogonal to all remaining
vectors in the list.
\begin{figure}[htbp]
\includegraphics{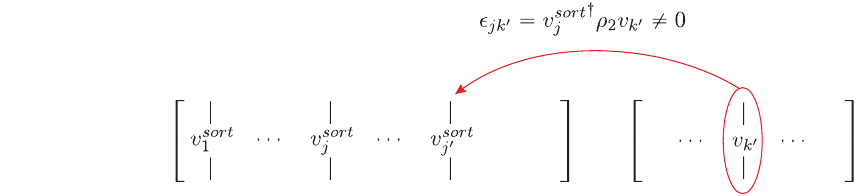}
\captionsetup{margin=0cm}
\caption{\hspace{2.75cm}{STEP 3.3}. Finding the remaining vectors in the same subspace as $v_1^{sort}$.}
\end{figure}

\item[3.4.] Once we have finished verifying all  eigenvectors in $L^{sort}$, we take a
block whose columns are the eigenvectors in $L^{sort}$ and denote it as $L^1$, since
it is a set of $c_1n_1$ vectors constituting an orthonormal basis of $\mathcal{L}^1$.
After that, we come back to STEP \hyperref[step_3_1]{3.1} and repeat the process with the rest of vectors
until all of them have been sorted.

\end{enumerate}

\medskip

At the end of this step, we obtain a matrix we may call $V_1^{sort_1}$ whose columns form bases $L^\alpha$ of the proper
subspaces $\mathcal{L}^\alpha$ for $\alpha=1,\ldots,N$, i.e.,
\[
\begin{array}{cccc}
  V_1^{sort_1} =\left(\begin{array}{c|c|c|c}
L^1\,\, & \,\,L^2\,\, & \,\cdots\,\, & \,\,L^N\end{array}\right).\vspace{-0.5cm} \\
\hspace{-1.2cm}\underbrace{\phantom{a..}}_{c_1n_1} & \hspace{-2.83cm}\underbrace{\phantom{aa}}_{c_2n_2} & \hspace{0.6cm}\underbrace{\phantom{aa.}}_{c_Nn_N}
 \end{array}
 \]
This step also gives the dimensions $c_\alpha n_\alpha$ by counting the number of vectors in each subspace.
\bigskip

\item[4.] \textbf{Second sorting}: Reorder the columns within each $L^{\alpha}$ grouping together the
eigenvectors corresponding to the same eigenvalue of $\rho_1$. To do this,
we just reorder the eigenvectors in each $L^\alpha$ in decreasing order corresponding to their eigenvalues. Thus, we obtain:
\begin{equation*}
V_1^{sort} =\left(\begin{array}{c|c|c|c}
{L^{1}}^{sort} & {L^{2}}^{sort} & \,\cdots\,\, & {L^{N}}^{sort} \end{array}\right),
\end{equation*}
where
$$
{{L^{\alpha}}^{sort}}^{\dagger}\rho_1{L^{\alpha}}^{sort}=\diag\left(\lambda_1^\alpha\mathds{1}_{c_\alpha},\lambda_2^\alpha\mathds{1}_{c_\alpha},\ldots,\lambda_{n_\alpha}^\alpha\mathds{1}_{c_\alpha}\right)\,.
$$

Counting the multiplicity of one eigenvalue in each $\alpha$ will give the multiplicity $c_\alpha$.  Hence, since we already got the
products $c_\alpha n_\alpha$ in STEP \hyperref[step_3]{3}, we can also get the dimensions of the irreps $n_\alpha$ by dividing those numbers by $c_\alpha$.
At this point, it is also possible, if needed, to obtain the characters of the irreps in the decomposition of $D(h)$ by computing
\begin{equation*}
\chi^\alpha(h)=\frac{1}{c_\alpha}\mathrm{Tr}\big({{L^{\alpha}}^{sort}}^{\dagger} D(h){L^{\alpha}}^{sort}\big)\,.
\end{equation*}

\item[5.] \textbf{Coarse block-diagonalization of $\boldsymbol{\rho_2}$}: Compute the matrix $V_1^{sort^{\dagger}}\rho_2\,V_1^{sort}$ to obtain the coarse block-diagonalization of $\rho_2$ in terms of the matrices $\Sigma^\alpha$, as shown in \textbf{Corollary \ref{Transformation_any_state}}, and identify  the square matrices $R_{ij},\ i,j=1,\ldots,n_\alpha$, of size $c_\alpha$.
\bigskip

\item[6.] \textbf{Compute a matrix $\widetilde{R}$}: According to \textbf{Corollary \ref{R_tilde:corollary}}, for each $\Sigma^\alpha$ choose a column of matrices $\widetilde{R}_{jk_\alpha}^\alpha$ such that $\widetilde{R}_{jk_\alpha}^\alpha\neq 0$ for all $j=1,\ldots,n_\alpha$, compute the unitary matrices
\begin{equation*}
\widetilde{R}_{k_\alpha}^\alpha=\diag\left(\widetilde{R}_{1k_\alpha}^\alpha,\widetilde{R}_{2k_\alpha}^\alpha,\ldots,\widetilde{R}_{n_\alpha k_\alpha}^\alpha\right)
\end{equation*}
 and finally compute the unitary matrix
\begin{equation*}
\widetilde{R}=\diag\left(\widetilde{R}_{k_1}^1,\widetilde{R}_{k_2}^2,\ldots,\widetilde{R}_{k_N}^N\right).
\end{equation*}

\smallskip

\item[7.] \textbf{Compute the permutation matrix $F$}: Matrices $F^{\alpha}$ will be the matrix $F$ in \textbf{Lemma \ref{permutation:lemma}} with $c=c_\alpha$ and $n=n_\alpha$, then compute those matrices for each $\alpha$ and collect them in the block diagonal matrix:
$$
\widetilde{F}=\diag\left(F^1,F^2,\ldots,F^N\right).
$$

\item[8.] \textbf{Final rearrangement}: Compute the Clebsch--Gordan matrix $\widehat{C}=V_{1}^{sort}\widetilde{R}\widetilde{F}$.
\end{enumerate}

\section{Some examples}\label{sec:examples}


\subsection{Decomposition of the regular representation of a finite group}\label{sec:regular}

The algorithm we have presented decomposes any finite-dimensional unitary representation of any  compact Lie group. In the case of finite groups, it is natural to apply it to the regular representation because it contains every irreducible representation with multiplicity equal to the dimension of its irreps, $c_\alpha=n_\alpha$ [\onlinecite{Se77}, Ch.\,2], thus: 
$$
|G|=\sum_{\alpha=1}^Nn_\alpha^2.
$$

The regular representation of a group $G$ is the unitary representation obtained from the action of the group $G$ on the Hilbert space of square integrable functions on the group, $\mathcal{H}=L^2(G,\mu)$, where $\mu$ denotes the left(right)-invariant Haar measure by left(right) translations.   

As before, we will restrict the discussion to finite groups $G$ as in \textbf{Sect.\,\ref{sec:preliminaries}}.
The space of square integrable functions on $G$ can be identified canonically with the $|G|$-dimensional complex space formally generated by the elements  of the group, i.e., we will denote by $\mathbb{C}[G]$ the linear space whose elements are given by $a = \sum_{g \in G} a_g g$, $a_g \in \mathbb{C}$, $g \in G$, with the natural addition law $a + b = \sum_{g \in G} (a_g + b_g) g$. Notice that $\mathbb{C}[G]$ carries also a natural associative algebra:
$$
a\cdot b = \sum_{g,g' \in G} a_g b_{g'} gg' = \sum_{g \in G} (\sum_{g'\in G} a_{gg'^{-1}}b_{g'}) g\,,
$$
although we will not make use of such structure here.

The left regular representation is defined as:
$$
U^{reg}(g) a = \sum_{g'\in G} a_{g'} gg'  = \sum_{g'\in G} a_{g^{-1}g'} g' \, .
$$

Thus, the matrix elements of the regular representation are obtained by computing the action of the group on the orthonormal basis $g_i$, $i=0,\ldots,n-1$, of the Hilbert space $\mathcal{H}=\mathbb{C}[G]$:
$$
D^{reg}_{ij}(g_k)=\langle g_i, U^{reg}(g_k)g_j\rangle=\langle g_i,g_kg_j\rangle.
$$
Then, the matrix representation of the left regular representation of the element $g_k$ can be easily computed from the table of the group written below (notice the inverse of the elements along the rows). The matrix $D^{reg}(g_k)$ is obtained by constructing a matrix with ones in the positions where $g_k$ appears in the table and zeros in the rest.

\begin{table}[htbp]
\centering
 \begin{tabular}{c|cccccc}
$
\begin{matrix}
\\
\vspace{-0.35cm}
\end{matrix}
$\textbf{T} &$e$&$g_1^{-1}$&\hspace{-0.1cm}$\cdots$ & $g_i^{-1}$ & \hspace{0.2cm}$\cdots$&$g_{n-1}^{-1}$\\\hline
$
\begin{matrix}
\\
\vspace{-0.3cm}
\end{matrix}
$
$e$&$e$&$g_1^{-1}$&\hspace{-0.1cm}$\cdots$ & $g_i^{-1}$ &\hspace{0.2cm}$\cdots$&$g_{n-1}^{-1}$\\
$g_1$&$g_1$&$e$&\hspace{-0.1cm}$\cdots$ & $g_1g_i^{-1}$ &\hspace{0.2cm}$\cdots$&$g_1g_{n-1}^{-1}$\\
$\vdots$&$\vdots$&$\vdots$&\hspace{-0.00cm}$\,\diagdots[-40]{1em}{.09em}$ &\vdots & &$\vdots$\\
$g_i$ & $g_i$ & $g_ig_1^{-1}$ &  & \!$e$ &  &  $g_ig_{n-1}^{-1}$ \\
$\vdots$&$\vdots$&$\vdots$& & \vdots &\hspace{0.0cm}$\!\diagdots[-40]{1em}{.09em}$&$\vdots$\\
$g_{n-1}$&$g_{n-1}$&$g_{n-1}g_1^{-1}$&\hspace{-0.1cm}$\cdots$ & $g_{n-1}g_i^{-1}$ &\hspace{0.2cm}$\cdots$&\;\;\,$e$\;\;.\\
 \end{tabular}
 \vspace{0.35cm}
 \caption{Group table.}\label{group_table}
 \end{table}

In the case of the regular representation, the input of our program can be the matrix $T$ constructed out of the table \textbf{T} (see TABLE \ref{group_table}) relabeled by identifying $e$ with $1$ and $g_i$ with $i+1$ and whose entries are defined as:
\begin{equation}
T_{ij}= k \, , \quad \mathrm{if} \quad  g_{i-1}g_{j-1}^{-1} = g_{k-1} \, , \qquad i,j,k = 1, \ldots , n \, .
\end{equation}
Once we have the group multiplication table in this form, we do not need to compute explicitly the regular representation for each element $D^{reg}(g)$ to create the adapted states $\rho_1$ and $\rho_2$ in STEP \hyperref[step_1]{1}, since we can simply evaluate the random vectors $\boldsymbol{\varphi}_a$ on the elements of the table, that is,
\begin{equation}
[\tau_a]_{ij}=\boldsymbol{\varphi}_a \big(T_{ij}\big)\, , \qquad a = 1,2 \, .
\end{equation}

In the final \hyperref[Appendix_ps]{\textbf{Appendix}}, we will show the results obtained using our algorithm for the decomposition of the regular representation in two simple cases: the permutation group $S_3$ and the alternating group $A_4$.

To verify the accuracy of the results, we will compare characters, since
they are independent of the choice of basis. We shall compute the characters $\widehat{\chi}^\alpha$
of the irreps obtained after applying the unitary transformation $\widehat{C}$
provided by our algorithm and we will compare them with the exact characters $\chi^\alpha_{exact}$ by defining the error as:
\begin{equation}\label{error_finite}
\widehat{\chi}_{error}=\frac{1}{|H|}\max_{\alpha\in\widehat{H}}\sum_{h\in H}|\chi^\alpha_{exact}(h)-\widehat{\chi}^\alpha(h)|\,,
\end{equation}
where  $\widehat{H}$ is the family of equivalence classes of irreps of $H$.

\subsection{Clebsch--Gordan coefficients of $SU(2)$}

Let $G$ be a compact Lie group and $H$ a closed subgroup (hence, compact too). States adapted to $H$ will have the form:
\begin{equation}\label{tomographic_decomposition_reducible_rep_2}
\rho=\frac{1}{Z}\int_H\overline{\chi_\rho(h)}D(h)\textrm{d} h\,,
\end{equation}
where $Z$ is the normalization factor
\begin{equation*}
Z=\int_H\overline{\chi_\rho(h)}\chi(h)\textrm{d}h\,,
\end{equation*}
and $\textrm{d}h$ denotes the invariant Haar measure on $H$.

Because our algorithm is numerical, we need to approximate the
integral (\ref{tomographic_decomposition_reducible_rep_2}) with a
finite sum. Choosing a quadrature rule to approximate the integral
(\ref{tomographic_decomposition_reducible_rep_2}) for a given
$\rho$ is equivalent to use another $\widehat{\rho}$ such that
$\chi_{\widehat{\rho}}\neq 0$ only at a finite number of elements of
the group. Then, the integral
(\ref{tomographic_decomposition_reducible_rep_2}) for $\widehat{\rho}$
reduces to a finite sum and the approximation of $\widehat{\rho}$ is
exact. It could happen that the generic adapted states thus
obtained do not have enough degrees of freedom, i.e.,
it might happen that the block diagonal matrices of the
representation were not irreducible. However, we will see that
this is not a problem because in the case of Lie groups, the
Clebsch--Gordan matrix  decomposing all the elements of its Lie
algebra $\mathfrak{g}$ will be the Clebsch--Gordan matrix
decomposing all the elements of the representation.

For compact Lie groups, the elements of a unitary representation are related via the exponential map with the corresponding representation via Hermitean matrices of elements of its Lie algebra $\mathfrak{g}$: $U(g)=\mathrm{e}^{is\xi}$, $s\in\mathds{R}$ and $\xi\in\mathfrak{g}$.

One can immediately see that the Clebsch--Gordan matrix $C$ that
decomposes the matrices representing all the elements of the Lie algebra
$\xi\in\mathfrak{g}$ will decompose all the elements of the unitary
representation and \textit{vice versa}:
$$
C^\dagger\xi_i C=\mathds{1}_{c_1}\otimes\xi_i^1\oplus\cdots\oplus\mathds{1}_{c_N}\otimes\xi_i^{N}\Longleftrightarrow\; C^\dagger U(g)C=\mathds{1}_{c_1}\otimes U^1(g)\oplus\cdots\oplus\mathds{1}_{c_N}\otimes U^N(g)\,,
$$
where $\{\xi_i^\alpha\}_{i=1}^{n_\mathfrak{g}}$,
$\alpha=1,\ldots,N$, are the matrices representing a set of generators of the Lie algebra $\mathfrak{g}$ ($n_\mathfrak{g}$ is the dimension of the set) and
$U^{\alpha}(g),\ \alpha=1,\ldots,N$, their corresponding unitary representations.

The original  Clebsch--Gordan problem consists in reducing a tensor
product representation $U_A(g) \otimes U_B(g)$, $\forall g\in G$,
of two representations of the same group $G$ restricted to the
diagonal subgroup of the product group. By associativity, this
problem can be generalized to any number of tensor product factors
$U^1(g) \otimes U^2(g)\otimes\cdots\otimes U^n(g)$. The associated Lie algebra generators will be given by:
\begin{equation*}
\xi_i=\xi_i^1\otimes\mathds{1}^2\otimes\cdots\otimes\mathds{1}^n+\mathds{1}^1\otimes\xi_i^2\otimes\cdots\otimes\mathds{1}^n+\cdots+\mathds{1}^1\otimes\mathds{1}^2\otimes\cdots\otimes \xi_i^n,
\end{equation*}
with commutation relations given by:
\begin{equation*}
[\xi_i,\xi_j]=c_{ij}^k\xi_k,\qquad[\xi_i^\alpha,\xi_i^\alpha]=c_{ij}^k\xi_i^\alpha,\qquad\alpha=1,\ldots,n,\qquad i,j,k=1,\ldots,n_\mathfrak{g},\qquad c_{ij}^k\in\mathds{C}.
\end{equation*}

Let us  now  study  the $SU(2)$ group: the generators of
the representation of its associated Lie algebra $\mathfrak{su}(2)$ are given by the
Hermitian traceless angular momentum operators $ J_k$ satisfying
the commutation relations
\begin{equation}\label{commutation_su2}
[J_i,J_j]=i\epsilon_{ij}^kJ_k\,, \qquad i,j,k=x,y,z\,,\qquad n_\mathfrak{g}=3\,.
\end{equation}
Its associated representation of $SU(2)$ can be written as:
\begin{equation}\label{exp_repsu2}
D(\boldsymbol{s})=\textrm{e}^{i\boldsymbol{s}\cdot\boldsymbol{J}}\,,\qquad \boldsymbol{s}=(s_x,s_y,s_z)\in\mathds{R}^3.
\end{equation}
The matrix representation of momentum $j$ of the angular momentum
operators $J_i$ is usually written in a basis of eigenvectors of
$J_z$,
$$
J_z|j,m\rangle=m|j,m\rangle\,,\qquad m=j,j-1,\ldots,-j\,,
$$
and the representation of the operators $J_x$ and $J_y$ is usually
obtained from the representation of the ladder operators
$J_{\pm}=J_x\pm iJ_y$,
\begin{equation}\label{ladder_operator}
\langle j,m|J_{\pm}|j,m'\rangle=\sqrt{(j\mp m')(j\pm m'+1)}\,\delta_{mm'\pm 1}\,.
\end{equation}
For instance, if $ j=3/2$:
$$
J_x=\begin{pmatrix}
0 &\frac{\sqrt{3}}{2}& 0& 0\\
\frac{\sqrt{3}}{2} &0 &1& 0\\
0&1& 0&\frac{\sqrt{3}}{2}\\
0 &0& \frac{\sqrt{3}}{2} & 0
\end{pmatrix}\,,\qquad J_y=\begin{pmatrix}
0 &-i\frac{\sqrt{3}}{2}& 0& 0\\
i\frac{\sqrt{3}}{2} &0 &-i& 0\\
0&i& 0&-i\frac{\sqrt{3}}{2}\\
0 &0& i\frac{\sqrt{3}}{2} & 0
\end{pmatrix},
$$
$$
J_z=\begin{pmatrix}
\frac{3}{2}&0&0&0\\
0&\frac{1}{2}&0&0\\
0&0&-\frac{1}{2}&0\\
0&0&0&-\frac{3}{2}
\end{pmatrix}\,,
$$
in the standard basis
$$
|3/2,3/2\rangle=\begin{pmatrix}
1\\
0\\
0\\
0
\end{pmatrix}\,,\quad|3/2,1/2\rangle=\begin{pmatrix}
0\\
1\\
0\\
0
\end{pmatrix}\,,\quad|3/2,-1/2\rangle=\begin{pmatrix}
0\\
0\\
1\\
0
\end{pmatrix}\,,\quad|3/2,-3/2\rangle=\begin{pmatrix}
0\\
0\\
0\\
1
\end{pmatrix}\,.
$$

The standard Clebsch--Gordan matrix is constructed with
eigenvectors of the total angular momentum operator $J_T$ with
respect to the $z$ component,
$$
{J_T}_z=J^1_z\otimes\mathds{1}^2\otimes\cdots\otimes\mathds{1}^n+\mathds{1}^1\otimes J^2_z\otimes\cdots\otimes\mathds{1}^n+\cdots+\mathds{1}^1\otimes\mathds{1}^2\otimes\cdots\otimes J^n_z\,,
$$
where $n$ is the number of parts of the system. The eigenvectors
of this  operator are usually denoted by $|J,M\rangle$, where $J$
represent the total angular momentum and $M=J,J-1,\ldots,-J$ the corresponding eigenvalues:
$$
{J_T}_z|J,M\rangle=M|J,M\rangle\,.
$$

The standard procedure to obtain this Clebsch--Gordan matrix consists in applying  successively the ladder operator $J_-$ starting from the state of maximum momentum $|J_{max},M_{max}\rangle=|j_1+j_2,j_1+j_2\rangle$. Notice that since the action of the matrix elements of the ladder operators~(\ref{ladder_operator}) is real, the Clebsh-Gordan coefficients are real too.

Recall that the Clebsch--Gordan matrix provided by our
algorithm is written in terms of the eigenvectors of the first
adapted state $\rho_1$. Thus, if we want to compare the
Clebsch--Gordan coefficients obtained from our algorithm with
the standard ones, we have to find a Clebsch--Gordan matrix $C_z$
which is conformed by eigenvectors of the operator ${J_T}_z$. To
do that, we first create two real adapted states using the fact
that the operators $J_k$ verify:
$$
{J}^*_x=J_x\,,\quad {J}^*_y=-J_y\,,\quad {J}^*_z=J_z\,,
$$
where $^*$ denotes the complex conjugate. Therefore, for  any adapted state $\rho$, its complex conjugate $\rho^*$ is an adapted state too. Hence, to create real adapted states, we first add to each matrix $\tau_a$, $a=1,2$, in STEP \hyperref[step_1]{1} in \textbf{Section \ref{the_algorithm:sec}}, its complex conjugate to obtain real symmetric matrices, and then we multiply the result by its transpose to make it positive definite. Finally, we normalize them, dividing by their trace, i.e.,
\begin{equation}\label{real_states}
\tilde{\rho}_a=\tau_a+\tau^*_a\,,\qquad {\rho_{real}}_a=\frac{1}{\mathrm{Tr}(\tilde{\rho}_a\tilde{\rho}_a^t)}\tilde{\rho}_a\tilde{\rho}_a^t.
\end{equation}

Once we have two real adapted states ${\rho_{real}}_1$ and
${\rho_{real}}_2$,  we apply our algorithm to get the real
Clebsch--Gordan matrix $\widehat{C}$. After that, we
transform the operator ${J_T}_z$ with $\widehat{C}$ to decompose
it into irreducible representations,
\begin{equation}\label{Jz_decomposition}
\widehat{C}^\dagger{J_T}_z \widehat{C}=\left(
\begin{array}{ccccc}
\begin{array}{|cc|}\hline
 \phantom{a}* &\hspace{0.3cm}  *\phantom{a}\\
  \phantom{a}* &\hspace{0.3cm}  *\phantom{a}\\\hline
\end{array}
 & &  &  &\\
 &\hspace{-0.21cm} \begin{array}{|cccc|}\hline
 \phantom{a}*&\hspace{0.3cm} * &\hspace{0.3cm} * &\hspace{0.3cm}  *\phantom{a}\\
 \phantom{a}*&\hspace{0.3cm}  *&\hspace{0.3cm}  *&\hspace{0.3cm} *\phantom{a} \\
  \phantom{a}*&\hspace{0.3cm} * &\hspace{0.3cm} * &\hspace{0.3cm} *\phantom{a} \\
  \phantom{a}*&\hspace{0.3cm} * &\hspace{0.3cm} * &\hspace{0.3cm} *\phantom{a}\\\hline
\end{array} &   & &\\
& &\hspace{-0.cm}\ddots & &\\
& & &\hspace{0.2cm}\ddots &\\
& & & &\hspace{0.05cm}\begin{array}{|ccc|}\hline
\phantom{a}* &\hspace{0.3cm} *&\hspace{0.3cm} *\phantom{a}\\
 \phantom{a}* &\hspace{0.3cm} *&\hspace{0.3cm} *\phantom{a}\\
 \phantom{a}* &\hspace{0.3cm} *&\hspace{0.3cm} *\phantom{a}\\\hline
\end{array}
 \end{array}
\right)\,,
\end{equation}
and  we  diagonalize each block of this matrix
transforming it with a block-diagonal matrix $V_z$ which reorders
the eigenvalues as follows:

 \begin{equation}\label{J_z_eigenvalues}
V_z^\dagger \widehat{C}^\dagger{J_T}_z \widehat{C}V_z=\begin{pmatrix}
j_1 & & & & & & & & &\\
& \hspace{-0cm}j_1-1 & & & & & & & &\\
& & \hspace{-0.7cm}\diagdots[-48]{1em}{.09em} & & & & & & & \\
& & & \hspace{-0.3cm}-j_1 & & & & & & \\
& & & & j_2 & & & & & \\
& & & & & \hspace{-0.2cm}j_2-1 & & & &\\
& & & & & &\hspace{-0.7cm}\diagdots[-48]{1em}{.09em} & & & \\
& & & & &  & & \hspace{-0.3cm}-j_2 & & \\
& & & & &  & & &\hspace{0.16cm}\diagdots[-48]{1em}{.09em}  & \\
& & & & &  & & & &\hspace{-2.4cm}\diagdots[-48]{1em}{.09em}  \\
& & & & &  & & & &\hspace{-1cm}j_N \\
& & & & &  & & & &\hspace{0.7cm}j_N-1 \\
& & & & &  & & & &\hspace{1.6cm}\diagdots[-48]{1em}{.09em}  \\
& & & & &  & & & &\hspace{2.5cm}-j_N \\
\end{pmatrix}\,.
\end{equation}
Therefore, the Clebsch--Gordan matrix whose columns are the
eigenvectors of ${J_T}_z$, reordered in this way, is given by
\begin{equation}
C_z=\widehat{C}V_z\,.
\end{equation}

In the \hyperref[Appendix_ps]{\textbf{Appendix}}, we will show the computation of the
Clebsch--Gordan  coefficients for the bipartite spin system
$3/2\times 1$ and for the tripartite spin system $1/2\times
1/2\times 3/2$. Again, we will verify the accuracy by comparing
the exact characters with the ones computed after transforming
with the Clebsch--Gordan matrix obtained with our algorithm. For
any irreducible representation of the $SU(2)$ group, it can be shown
that the characters have the following expression:
\begin{equation}
\chi^n_{exact}(\boldsymbol{s})=\begin{cases}
\displaystyle{2\sum\limits_{k=1}^{n/2}\cos\left(\sqrt{s_x^2+s_y^2+s_z^2}\left(\frac{n-1}{2}-k+1\right)\right)}\,,&\text{$n$ even,}\\
\displaystyle{2\hspace{-0.27cm}\sum\limits_{k=1}^{(n-1)/2}\hspace{-0.27cm}\cos\left(\sqrt{s_x^2+s_y^2+s_z^2}\left(\frac{n-1}{2}-k+1\right)\right)+1}\,,&\text{$n$ odd},
\end{cases}
\end{equation}
where $n=2j+1$ is the dimension of the irrep. Therefore, we measure the accuracy through
\begin{equation}\label{error_Lie}
\widehat{\chi}_{error}=\max_{\alpha\in\widehat{H}}\int_{H}|\chi^\alpha_{exact}(h)-\widehat{\chi}^\alpha(h)|\textrm{d}h\approx\frac{1}{N_H}\sum_{i=1}^{N_H}|\chi^\alpha_{exact}(h)-\widehat{\chi}^\alpha(h)|\,,
\end{equation}
with $N_H$ the number of elements in the quadrature approximation.


\section{Conclusions and discussion}\label{sec:discussion}

A numerical algorithm to compute the decomposition of a
finite-dimensional unitary representation of a compact Lie group has
been presented. Such algorithm uses the notion of generic
adapted quantum mixed states to obtain the block structure and,
eventually, the coefficients of the Clebsch--Gordan matrix solving
the decomposition problem.

The numerical algorithm is stable and accurate, since it combines nothing but
stable routines involving diagonalization of Hermitian matrices,
sorting and recombination of matrix blocks and matrix products.
The numerical examples presented confirm this.

The algorithm has been used successfully to decompose the
regular representation of two finite groups and the direct product of
two and three representations of $SU(2)$. In the first case, the main
computational task was to prepare the group table, a preliminary
task before the algorithm is used. In the second case, this preliminary
part was much easier, since explicit expressions of the representations
of the Lie algebra $\mathfrak{su}(2)$, for any value of spin, are well-known.

The algorithm can be easily extended to finite-dimensional
representations of non-compact groups. However, because the
representations will cease to be unitary, the numerical  stability
of the algorithm could be compromised.    Further insights on
these questions will be considered elsewhere.

\phantomsection\label{Appendix_ps}\section*{Appendix}

In this appendix, we present the results obtained for the
decomposition of  the $S_3$ and $A_4$ group, and the
Clebsch--Gordan coefficients of the spin systems $3/2\times 1$ and
$1/2\times 1/2\times 3/2$. All experiments were conducted
using Matlab R2012a (version 7.14.0.739).

\bigskip

\subsection*{\texorpdfstring{\noindent \textbf{A.1.}\;\,\textbf{The decomposition of the left regular
representation of the permutation group $S_3$.}}{The decomposition of the left regular
representation of the permutation group $S_3$}} The $S_3$ group is
the group of permutations of three elements and it has order six.
The elements of this group can be generated with the set of
transpositions $a_k=(k,k+1)$, $k=1,2$:
 \begin{equation*}
 a_1^2=a_2^2=(a_1a_2)^3=e\,.
 \end{equation*}
Our algorithm decomposes the regular representation into two representations
$\widehat{D}^1$ and $\widehat{D}^2$ of dimension one and multiplicity one,
and another one $\widehat{D}^3$ of dimension two and multiplicity two, exactly
as expected. The representation $\widehat{D}^1$ corresponds to the trivial
one, $\widehat{D}^1(g)=1$, $\forall g\in S_3$, and the rest of representations
obtained after applying the transformation $\widehat{C}$ are the following:
\begin{table*}[htbp]
$$
\begin{array}{c!{\vrule width 1.5pt}c!{\vrule width 1pt}c|}
S_3&\widehat{D}^2&\widehat{D}^3\\\noalign{\hrule height 1.5pt}
\begin{matrix}
\phantom{a}\\
e\\
\phantom{a}
\end{matrix}&\begin{matrix}
\phantom{a}\\
1.0000\\
\phantom{a}
\end{matrix}&\begin{pmatrix}
1.0000&0.0000+0.0000i\\
0.0000-0.0000i&1.0000
\end{pmatrix}\\\noalign{\hrule}
\begin{matrix}
\phantom{a}\\
a_1\\
\phantom{a}
\end{matrix}&\begin{matrix}
\phantom{a}\\
-1.0000\\
\phantom{a}
\end{matrix}&\begin{pmatrix}
-0.7501&0.6399-0.1671i\\
0.6399+0.1671i&0.7501
\end{pmatrix}\\\noalign{\hrule}
\end{array}
$$
\end{table*}
\\
\begin{table}[htbp]
$$
\begin{array}{c!{\vrule width 1.5pt}c!{\vrule width 1pt}c|}
S_3&\widehat{D}^2&\widehat{D}^3\\\noalign{\hrule height 1.5pt}
\begin{matrix}
\phantom{a}\\
a_2\\
\phantom{a}
\end{matrix}&\begin{matrix}
\phantom{a}\\
-1.0000\\
\phantom{a}
\end{matrix}&\begin{pmatrix}
0.3542&-0.5615-0.7479i\\
-0.5615+0.7479i&-0.3542
\end{pmatrix}\\\noalign{\hrule}
\begin{matrix}
\phantom{a}\\
a_1a_2\\
\phantom{a}
\end{matrix}&\begin{matrix}
\phantom{a}\\
1.0000\\
\phantom{a}
\end{matrix}&\begin{pmatrix}
-0.5000+0.5723i&0.1945+0.6202i\\
-0.1945+0.6202i&-0.5000-0.5723i
\end{pmatrix}\\\noalign{\hrule}
\begin{matrix}
\phantom{a}\\
a_2a_1\\
\phantom{a}
\end{matrix}&\begin{matrix}
\phantom{a}\\
1.0000\\
\phantom{a}
\end{matrix}&\begin{pmatrix}
-0.5000-0.5723i&-0.1945-0.6202i\\
0.1945-0.6202i&-0.5000+0.5723i
\end{pmatrix}\\\noalign{\hrule}
\begin{matrix}
\phantom{a}\\
a_2a_1a_2\\
\phantom{a}
\end{matrix}&\begin{matrix}
\phantom{a}\\
-1.0000\\
\phantom{a}
\end{matrix}&\begin{pmatrix}
0.3959&-0.0784+0.9149i\\
-0.0784-0.9149i&-0.3959
\end{pmatrix}\\\noalign{\hrule}
\end{array}
$$
\vspace{0.1cm}
\caption{Irreducible representations obtained for $S_3$ group.}
\end{table}

If we use the formula \eqref{error_finite} to compute the accuracy of the characters of the irreps, we obtain:
$$
\widehat{\chi}_{error}=3.5785\cdot 10^{-15}\,.
$$

 \medskip

\subsection*{\texorpdfstring{\noindent \textbf{A.2.}\;\,\textbf{The decomposition of the left regular representation of the alternating group $A_4$.}}{The decomposition of the left regular representation of the alternating group $A_4$}} The alternating group $A_4$ is the group of even permutations of four elements. This group has twelve elements and it can be generated with three generators satisfying the relations
\begin{equation*}\label{A4_relations}
a^2=b^2=c^3=(ab)^2=ac^2abc=bc^2ac=e.
\end{equation*}
The left regular representation of this group has four irreducible representations: three of dimension one and one of dimension three. Hence, our algorithm will decompose the regular representation of this group into the three representations of dimension one with multiplicity one and the representation of dimension three with multiplicity three. Again, $\widehat{D}^1$ is the trivial representation $\widehat{D}^1(g)=1$, $\forall g\in A_4$, and the rest are given by:
\begin{table*}[htbp]
$$
\begin{array}{c!{\vrule width 1.5pt}c!{\vrule width 1pt}c!{\vrule width 1pt}c|}
A_4\;\;& \widehat{D}^2&\widehat{D}^3&\widehat{D}^4\\\noalign{\hrule height 1.5pt}
\begin{matrix}
\vspace{0.2cm}\phantom{a}\\
e\\
\vspace{0.2cm}\phantom{a}\end{matrix}&\hspace{-0.05cm}\begin{matrix}
\vspace{0.2cm}\phantom{a}\\
1.0000\\
\vspace{0.2cm}\phantom{a}\end{matrix}\hspace{-0.2cm}&\hspace{-0.05cm}\begin{matrix}
\vspace{0.2cm}\phantom{a}\\
1.0000\\
\vspace{0.2cm}\phantom{a}\end{matrix}\hspace{-0.3cm}&\hspace{0.cm}\begin{pmatrix}
1.0000    &        -0.0000 + 0.0000i & -0.0000 - 0.0000i\\
  -0.0000 - 0.0000i &  1.0000  &           0.0000 + 0.0000i\\
  -0.0000 + 0.0000i &  0.0000 - 0.0000i &  1.0000
\end{pmatrix}\hspace{-0.3cm}\\\noalign{\hrule}
\begin{matrix}
\vspace{0.2cm}\phantom{a}\\
a\\
\vspace{0.2cm}\phantom{a}\end{matrix}&\hspace{-0.05cm}\begin{matrix}
\vspace{0.2cm}\phantom{a}\\
1.0000\\
\vspace{0.2cm}\phantom{a}\end{matrix}\hspace{-0.2cm}&\hspace{-0.05cm}\begin{matrix}
\vspace{0.2cm}\phantom{a}\\
1.0000\\
\vspace{0.2cm}\phantom{a}\end{matrix}\hspace{-0.3cm}&\hspace{-0.1cm}\begin{pmatrix}
-0.9852     &       -0.0240 + 0.0941i &  0.1176 + 0.0789i\\
  -0.0240 - 0.0941i & -0.3653       &      0.3099 - 0.8724i\\
   0.1176 - 0.0789i &  0.3099 + 0.8724i &  0.3504
\end{pmatrix}\hspace{-0.3cm}\\\noalign{\hrule}
\begin{matrix}
\vspace{0.2cm}\phantom{a}\\
b\\
\vspace{0.2cm}\phantom{a}\end{matrix}&\hspace{-0.05cm}\begin{matrix}
\vspace{0.2cm}\phantom{a}\\
1.0000\\
\vspace{0.2cm}\phantom{a}\end{matrix}\hspace{-0.2cm}&\hspace{-0.05cm}\begin{matrix}
\vspace{0.2cm}\phantom{a}\\
1.0000\\
\vspace{0.2cm}\phantom{a}\end{matrix}\hspace{-0.3cm}&\hspace{0.cm}\begin{pmatrix}
0.6482       &     -0.2501 + 0.4766i&  -0.3940 - 0.3672i\\
  -0.2501 - 0.4766i&  -0.8242       &     -0.0464 + 0.1697i\\
  -0.3940 + 0.3672i & -0.0464 - 0.1697i & -0.8240 - 0.0000i
\end{pmatrix}\hspace{-0.3cm}\\\noalign{\hrule}
\begin{matrix}
\vspace{0.2cm}\phantom{a}\\
c\\
\vspace{0.2cm}\phantom{a}\end{matrix}&\hspace{-0.05cm}\begin{matrix}
\vspace{-0.2cm}\phantom{a}\\
\!\!-0.5000\\
\,+0.866i\\
\vspace{-0.2cm}\phantom{a}\end{matrix}\hspace{-0.2cm}&\hspace{-0.15cm}\begin{matrix}
\vspace{-0.2cm}\phantom{a}\\
\!\!-0.5000\\
\,\,-0.8660i\\
\vspace{-0.2cm}\phantom{a}\end{matrix}\hspace{-0.3cm}&\hspace{-0.1cm}\begin{pmatrix}
-0.1137 - 0.4209i & -0.4113 - 0.2302i &  0.4649 - 0.6096i\\
  -0.0136 + 0.5419i &  0.0028 + 0.5742i  & 0.5988 - 0.1335i\\
  -0.6284 + 0.3482i  & 0.4483 - 0.4971i &  0.1110 - 0.1533i
\end{pmatrix}\hspace{-0.3cm}\\\noalign{\hrule}
\begin{matrix}
\vspace{0.2cm}\phantom{a}\\
c^2\\
\vspace{0.2cm}\phantom{a}\end{matrix}&\hspace{-0.15cm}\begin{matrix}
\vspace{-0.2cm}\phantom{a}\\
\!\!-0.5000\\
\,\,-0.8660i\\
\vspace{-0.2cm}\phantom{a}\end{matrix}\hspace{-0.5cm}&\hspace{-0.15cm}\begin{matrix}
\vspace{-0.2cm}\phantom{a}\\
\!\!-0.5000\\
\,+0.8660i\\
\vspace{-0.2cm}\phantom{a}\end{matrix}\hspace{-0.3cm}&\hspace{0.cm}\begin{pmatrix}
-0.1137 + 0.4209i & -0.0136 - 0.5419i & -0.6284 - 0.3482i\\
  -0.4113 + 0.2302i  & 0.0028 - 0.5742i  & 0.4483 + 0.4971i\\
   0.4649 + 0.6096i  & 0.5988 + 0.1335i &  0.1110 + 0.1533i
   \end{pmatrix}\hspace{-0.3cm}\\\noalign{\hrule}
\begin{matrix}
\vspace{0.2cm}\hspace{-0.1cm}\phantom{a}\\
ab\\
\vspace{0.2cm}\phantom{a}\end{matrix}&\hspace{-0.15cm}\begin{matrix}
\vspace{0.2cm}\phantom{a}\\
1.0000\\
\vspace{0.2cm}\phantom{a}\end{matrix}\hspace{-0.3cm}&\hspace{-0.05cm}\begin{matrix}
\vspace{0.2cm}\phantom{a}\\
1.0000\\
\vspace{0.2cm}\phantom{a}\end{matrix}\hspace{-0.3cm}&\hspace{-0.1cm}\begin{pmatrix}
-0.6631      &       0.2741 - 0.5707i &  0.2765 + 0.2883i\\
   0.2741 + 0.5707i &  0.1895       &     -0.2635 + 0.7028i\\
   0.2765 - 0.2883i & -0.2635 - 0.7028i & -0.5264
\end{pmatrix}\hspace{-0.3cm}\\\noalign{\hrule}
\begin{matrix}
\vspace{0.2cm}\phantom{a}\\
cb\\
\vspace{0.2cm}\phantom{a}\end{matrix}&\hspace{-0.15cm}\begin{matrix}
\vspace{-0.2cm}\phantom{a}\\
\!-0.5000\\
\,+0.8660i\\
\vspace{-0.2cm}\phantom{a}\end{matrix}\hspace{-0.3cm}&\hspace{-0.15cm}\begin{matrix}
\vspace{-0.2cm}\phantom{a}\\
\!\!-0.5000\\
\,\,-0.8660i\\
\vspace{-0.2cm}\phantom{a}\end{matrix}\hspace{-0.3cm}&\hspace{0.cm}\begin{pmatrix}
-0.0400 + 0.3917i &  0.4431 + 0.1902i & -0.4347 + 0.6508i\\
   0.0772 + 0.4789i & -0.3076 - 0.7107i & -0.3866 - 0.1247i\\
  -0.7438 + 0.2375i & -0.4095 + 0.0115i &  0.3475 + 0.3190i
\end{pmatrix}\hspace{-0.3cm}\\\noalign{\hrule}
\end{array}
$$
\end{table*}
\begin{table}[htbp]
$$
\begin{array}{c!{\vrule width 1.5pt}c!{\vrule width 1pt}c!{\vrule width 1pt}c|}
\noalign{\hrule}
\begin{matrix}
\vspace{0.2cm}\phantom{a}\\
ca\\
\vspace{0.2cm}\phantom{a}\end{matrix}&\hspace{-0.15cm}\begin{matrix}
\vspace{-0.2cm}\phantom{a}\\
\!-0.5000\\
\,+0.8660i\\
\vspace{-0.2cm}\phantom{a}\end{matrix}\hspace{-0.3cm}&\hspace{-0.15cm}\begin{matrix}
\vspace{-0.2cm}\phantom{a}\\
\!\!-0.5000\\
\,\,-0.8660i\\
\vspace{-0.2cm}\phantom{a}\end{matrix}\hspace{-0.3cm}&\hspace{-0.1cm}\begin{pmatrix}
0.1069 + 0.3505i  & 0.8684 + 0.3002i & -0.1455 + 0.0155i\\
   0.1273 - 0.6109i &  0.2504 + 0.2570i  & 0.6673 + 0.1914i\\
   0.5625 - 0.4001i & -0.0133 + 0.1634i & -0.3573 - 0.6075i
\end{pmatrix}\hspace{-0.3cm}\\\noalign{\hrule}
\begin{matrix}
\vspace{0.2cm}\phantom{a}\\
bc\\
\vspace{0.2cm}\phantom{a}\end{matrix}&\hspace{-0.05cm}\begin{matrix}
\vspace{-0.2cm}\phantom{a}\\
\!\!-0.5000\\
\,+0.866i\\
\vspace{-0.2cm}\phantom{a}\end{matrix}\hspace{-0.2cm}&\hspace{-0.15cm}\begin{matrix}
\vspace{-0.2cm}\phantom{a}\\
\!\!-0.5000\\
\,\,-0.8660i\\
\vspace{-0.2cm}\phantom{a}\end{matrix}\hspace{-0.3cm}&\hspace{0.cm}\begin{pmatrix}
0.0468 - 0.3213i & -0.9002 - 0.2602i  & 0.1153 - 0.0567i\\
  -0.1908 - 0.4100i &  0.0544 - 0.1205i & -0.8795 + 0.0669i\\
   0.8097 - 0.1857i & -0.0255 + 0.3222i & -0.1013 + 0.4419i
\end{pmatrix}\hspace{-0.3cm}\\\noalign{\hrule}
\begin{matrix}
\vspace{0.2cm}\phantom{a}\\
bc^2\\
\vspace{0.2cm}\phantom{a}\end{matrix}&\hspace{-0.15cm}\begin{matrix}
\vspace{-0.2cm}\phantom{a}\\
\!\!-0.5000\\
\,\,-0.8660i\\
\vspace{-0.2cm}\phantom{a}\end{matrix}\hspace{-0.5cm}&\hspace{-0.15cm}\begin{matrix}
\vspace{-0.2cm}\phantom{a}\\
\!\!-0.5000\\
\,+0.8660i\\
\vspace{-0.2cm}\phantom{a}\end{matrix}\hspace{-0.3cm}&\hspace{0.cm}\begin{pmatrix}
-0.0400 - 0.3917i &  0.0772 - 0.4789i & -0.7438 - 0.2375i\\
   0.4431 - 0.1902i & -0.3076 + 0.7107i & -0.4095 - 0.0115i\\
  -0.4347 - 0.6508i & -0.3866 + 0.1247i &  0.3475 - 0.3190i
\end{pmatrix}\hspace{-0.3cm}\\\noalign{\hrule}
\begin{matrix}
\vspace{0.2cm}\phantom{a}\\
cbc\\
\vspace{0.2cm}\phantom{a}\end{matrix}&\hspace{-0.15cm}\begin{matrix}
\vspace{-0.2cm}\phantom{a}\\
\!\!-0.5000\\
\,\,-0.8660i\\
\vspace{-0.2cm}\phantom{a}\end{matrix}\hspace{-0.5cm}&\hspace{-0.15cm}\begin{matrix}
\vspace{-0.2cm}\phantom{a}\\
\!\!-0.5000\\
\,+0.8660i\\
\vspace{-0.2cm}\phantom{a}\end{matrix}\hspace{-0.3cm}&\hspace{-0.1cm}\begin{pmatrix}
0.1069 - 0.3505i &  0.1273 + 0.6109i  & 0.5625 + 0.4001i\\
   0.8684 - 0.3002i &  0.2504 - 0.2570i & -0.0133 - 0.1634i\\
  -0.1455 - 0.0155i  & 0.6673 - 0.1914i & -0.3573 + 0.6075i
\end{pmatrix}\hspace{-0.3cm}\\\noalign{\hrule}
\begin{matrix}
\vspace{0.2cm}\phantom{a}\\
c^2b\\
\vspace{0.2cm}\phantom{a}\end{matrix}&\hspace{-0.15cm}\begin{matrix}
\vspace{-0.2cm}\phantom{a}\\
\!\!-0.5000\\
\,\,-0.8660i\\
\vspace{-0.2cm}\phantom{a}\end{matrix}\hspace{-0.5cm}&\hspace{-0.15cm}\begin{matrix}
\vspace{-0.2cm}\phantom{a}\\
\!\!-0.5000\\
\,+0.8660i\\
\vspace{-0.2cm}\phantom{a}\end{matrix}\hspace{-0.3cm}&\hspace{0.cm}\begin{pmatrix}
0.0468 + 0.3213i & -0.1908 + 0.4100i &  0.8097 + 0.1857i\\
  -0.9002 + 0.2602i &  0.0544 + 0.1205i  &-0.0255 - 0.3222i\\
   0.1153 + 0.0567i & -0.8795 - 0.0669i & -0.1013 - 0.4419i
\end{pmatrix}\hspace{-0.3cm}\\\noalign{\hrule}
\end{array}
$$
\vspace{0.1cm}
\caption{Irreducible representations obtained for $A_4$ group.}
\end{table}

In this case, the accuracy of the characters of the irreps computed with (\ref{error_finite}) is given by 
$$
\widehat{\chi}_{error}=4.4888\cdot 10^{-15}.
$$

\subsection*{\texorpdfstring{\noindent \textbf{B.1.}\;\,\textbf{Clebsch--Gordan coefficients for the spin system $3/2\times 1$.}}{Clebsch--Gordan coefficients for the spin system $3/2\,\textrm{x}\,1$}} Suppose we have a system of two particles in which the first particle has momentum $3/2$ and the second, momentum $1$. It is well known [\onlinecite{Ga90}, Ch.\,5] that this system is decomposed in the direct sum of systems of momentum $5/2$, $3/2$ and $1/2$, each one with multiplicity one: 
$$
3/2\times 1=5/2\oplus 3/2\oplus 1/2,
$$
or, in other words, that the representation of $SU(2)$ corresponding to the tensor product $3/2\times 1$ has irreducible representations with momentum $5/2$, $3/2$ and $1/2$ with multiplicity one each other.

To create the adapted states for STEP \hyperref[step_1]{1} of the algorithm, we have chosen three random vectors $\boldsymbol{s}_i=({s_x}_i,{s_y}_i,{s_z}_i)$, $\boldsymbol{s}_i\neq\textbf{0}$, $i=1,2,3$, for each adapted state, to obtain the three linearly independent elements of the representation. Obviously, we have also created two random vectors $\boldsymbol{\varphi}_a$ of length $3$ to construct the matrices $\tau_a$, $a=1,2$ in STEP \hyperref[step_1]{1}:
$$
\tau_a=\mathds{1}+\sum_{i=1}^3{\varphi_a}_iD^{3/2}({\boldsymbol{s}_a}_i)\otimes D^{1}({\boldsymbol{s}_a}_i)\,,
$$
where $D^{j_\alpha}(\boldsymbol{s})$ is the exponential representation given by (\ref{exp_repsu2}) and  $j_\alpha$ denotes the momentum of the representation $\alpha$.

To represent the computed Clebsch--Gordan coefficients, we will use the
following standard arrangement:
\bigskip
\begin{center}
\begin{figure*}[htbp]
\centering
\includegraphics{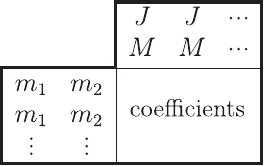}
\end{figure*}
\end{center}

The coefficients obtained for the system $3/2\times 1$ applying the algorithm are shown in the following table:
\bigskip
\medskip
\begin{table*}[htbp]
\hspace{-3.5cm}\includegraphics{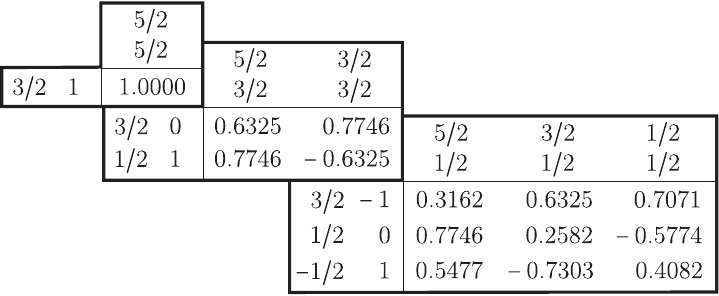}
\end{table*}
\newpage
\vspace{-0.8cm}\begin{table}[htbp]
\hspace{3.8cm}\includegraphics{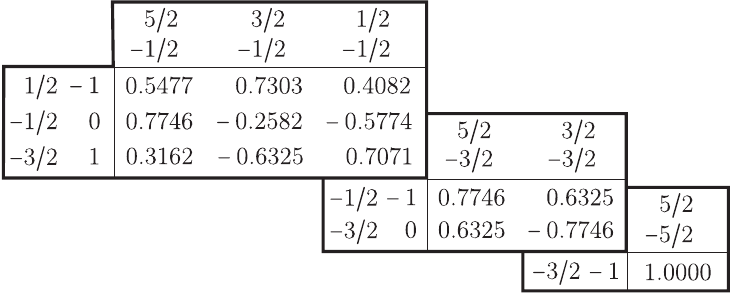}
\vspace{0.5cm}
\caption{CG coefficients for $3/2\times 1$.}
\end{table}

To assess the accuracy, we have approximated the integral in (\ref{error_Lie}) with $N_H=50^3$.
The result we obtained is:
$$
\widehat{\chi}_{error}=2.2340\cdot10^{-16}\,.
$$

\subsection*{\texorpdfstring{\noindent \textbf{B.2.}\;\,\textbf{Clebsch--Gordan coefficients for the
spin system $1/2\times 1/2\times 3/2$.}}{Clebsch--Gordan coefficients for the
spin system $1/2\,\textrm{x}\,1/2\,\textrm{x}\,3/2$}} To test the capabilities
of our algorithm, we will compute the Clebsch--Gordan coefficients
of a system of three spin particles. These coefficients can be
obtained from suitable choices of coefficients of products of two
spins, for that reason, there are no exhaustive tables for systems with more than
two particles.

The standard procedure consists in first reducing the representation
of the first two particles, then reducing the result with the next
particle, and so on, until there are no particles left. In our case,
the product of three particles with spin $1/2$, $1/2$ and $3/2$ yields:
\begin{equation*}\label{product_3_spins}
1/2\otimes 1/2\otimes 3/2=(0\oplus 1)\otimes 3/2=3/2\oplus 5/2\oplus 3/2\oplus 1/2,
\end{equation*}
this is, two irreps of momentum $1/2$ and $5/2$ with multiplicity one and other of momentum $3/2$ with multiplicity two.

In the first step, we block-diagonalize the first two spins:
\begin{equation*}
(C_{1/2\otimes1/2}\otimes\mathds{1}_4)^\dagger(D^{1/2}\otimes D^{1/2}\otimes D^{3/2})(C_{1/2\otimes1/2}\otimes\mathds{1}_4)
=\big(D^{0}\oplus D^{1}\big)\otimes D^{3/2}
\end{equation*}
and then, we diagonalize the result:
\begin{equation*}
\begin{pmatrix}
                   \mathds{1}_4 &0  \\
                    0 &C_{1\otimes 3/2}^\dagger
\end{pmatrix}\left((D^{0}\oplus D^{1})\otimes D^{3/2}\right)\begin{pmatrix}
                   \mathds{1}_4 &0  \\
                    0 &C_{1\otimes 3/2}
\end{pmatrix}=D^{3/2}\oplus D^{5/2}\oplus D^{3/2}\oplus D^{1/2}.
\end{equation*}
Therefore, the Clebsch--Gordan matrix of this system is
\begin{equation*}\label{Clebsh_three_spins}
C_{1/2\otimes 1/2\otimes 3/2}=(C_{1/2\otimes 1/2}\otimes\mathds{1}_4)(\mathds{1}_4\oplus C_{1\otimes 3/2}).
\end{equation*}

In this example, we see that for a multipartite system of spins,
the multiplicities of the representations can be bigger than one.
Thus, several eigenvectors may exist with the same values of $J$ and $M$.
Therefore, it is necessary to add another `quantum number', which
we will denote by $c$, to tell them apart. This `quantum number' will be
a label indicating to which copy of the representation of multiplicity
larger than one  each of the eigenvectors with the same $J$ and $M$ belongs
(for that reason, the choice of $c$ to denote it, since this is the letter we used to denote
 multiplicities in  (\ref{decompositionU}) above).
 
Using our algorithm, we do not need to group the system into groups of bipartite
systems as before and the computation can be done in one step. Again, in this case, we have chosen
three random vectors $\boldsymbol{s}_i$, $i=1,2,3$, to obtain three linearly
independent elements of the representation of the group, and two random
vector $\boldsymbol{\varphi}_a$ of length $3$ to compute the linear combinations
$\tau_a$, $a=1,2$. The coefficients will be represented in  arrangements similar to
the case of two spins but now including the label $c$:
\bigskip
\begin{center}
\begin{figure*}[htbp]
\centering
\includegraphics{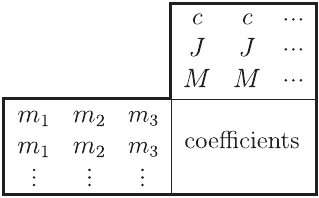}
\end{figure*}
\end{center}

Notice that the TABLE \hyperref[Tabla_5]{\ref*{table_3_spins}} below, corresponding to the Clebsch--Gordan coefficients of the tripartite system $1/2\times 1/2\times 3/2$, is not unique because there exists more than one linear combination providing a valid Clebsch--Gordan matrix that diagonalizes ${J_T}_{z}$ with the eigenvalues reordered in the way given in (\ref{J_z_eigenvalues}). 
\newpage
\phantomsection\label{Tabla_5}
\begin{table}[htbp]
\centering
\includegraphics{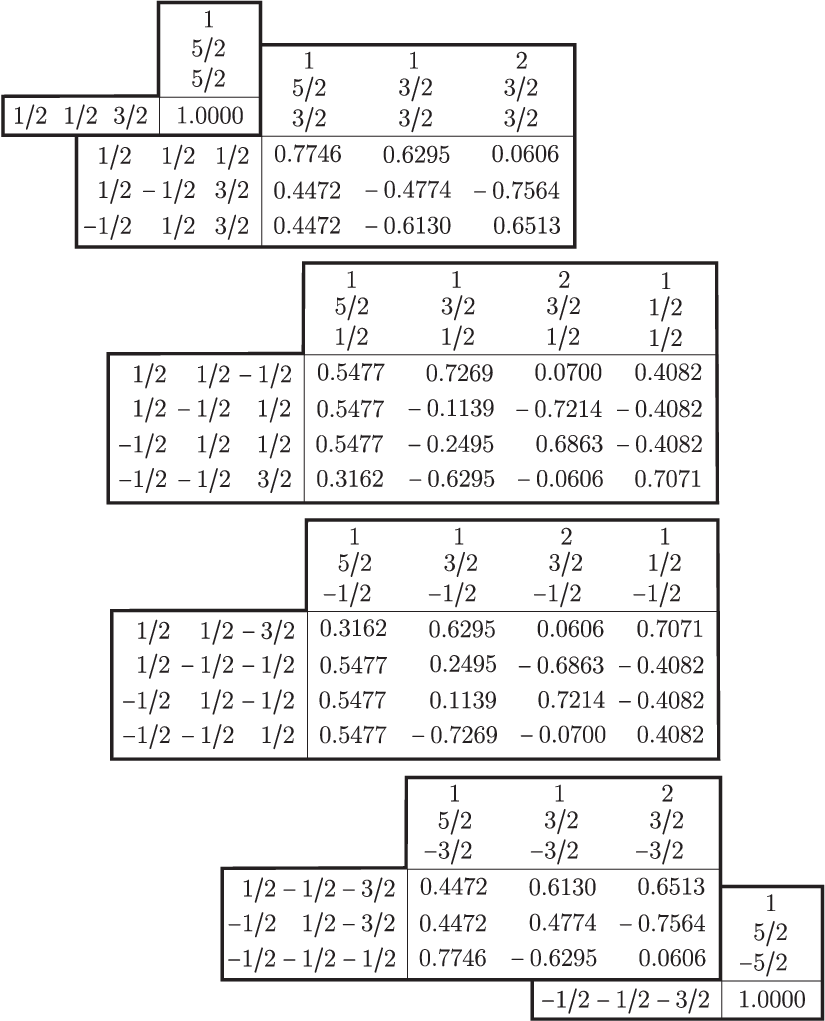}
\vspace{0.5cm}
\caption{CG coefficients for $1/2\times 1/2\times 3/2$.}
\label{table_3_spins}
\end{table}

Again, to assess the accuracy, we have approximated the integral in \eqref{error_Lie} with $N_H=50^3$, and the result obtained was
$$
\widehat{\chi}_{error}=5.2888\cdot10^{-15}\,.
$$

\end{document}